\pdfoutput=1

\documentclass[reprint,aps,prb,twocolumn,floatfix,superscriptaddress]{revtex4-2}
\usepackage[linktocpage=true,colorlinks=true,urlcolor=blue,linkcolor=red,citecolor = blue]{hyperref}
\usepackage{graphicx}
\usepackage{amssymb}
\usepackage{bm}
\usepackage{amsmath}
\usepackage{dsfont}
\usepackage{lineno}
\usepackage[english]{babel}
\usepackage[utf8]{inputenc}
\usepackage{xcolor}
\RequirePackage[T1]{fontenc}
\RequirePackage{lmodern}
\usepackage{appendix}
\usepackage{float}
\usepackage{algpseudocode}
\usepackage{algorithm}
\usepackage{mathtools}
\usepackage{filecontents}
\usepackage{physics}
\usepackage{comment}
\usepackage{siunitx}
\usepackage{bbm}
\usepackage{siunitx}
\usepackage{url}
\usepackage{verbatim}
\usepackage{subfiles} % Best loaded last in the preamble

\begin{document}

\def\E{\mathbb E}
\def\H{\mathcal H}
\def\N{\mathcal N}
\def\P{\mathbb P}
\def\Q{\mathbb Q}
\def\R{\mathbb R}
\def\bfp{\mathbf p}
\def\bfq{\mathbf q}
\def\bfx{\mathbf x}
\def\V{\mathbb V}
\def\argmax{\mathrm{argmax}}
\def\argmin{\mathrm{argmin}}
\def\sgn{\mathrm{sgn}}

%TC:ignore

\title{Using a spin-triplet encoding to enhance shuttling fidelities in Si/SiGe quantum wells}

\author{Merritt P. R. Losert}
\altaffiliation{merritt.losert@nist.gov}
\affiliation{University of Wisconsin-Madison, Madison, WI 53706 USA}
\affiliation{National Institute of Standards and Technology, Gaithersburg, Maryland 20899, USA}
\affiliation{Joint Center for Quantum Information and Computer Science, University of Maryland, College Park, Maryland 20742, USA}

\author{S. N. \surname{Coppersmith}}
\affiliation{School of Physics, University of New South Wales, Sydney, New South Wales 2052, Australia}

\author{Mark Friesen}
\affiliation{University of Wisconsin-Madison, Madison, WI 53706 USA}

\date{\today}
\pacs{}

\begin{abstract} % abstract
Spatial variations of the valley splitting in a quantum well present a key challenge for conveyor-mode shuttling of electron spins in Si/SiGe, giving rise to Landau-Zener-like excitations that cause leakage outside the qubit subspace. 
Here, we propose an unconventional two-electron qubit encoding, based on valley-singlet states, that is largely immune to Landau-Zener leakage processes. 
In contrast to single-electron spins, the shuttling fidelity actually \emph{improves} for small valley splittings, in this case.
We show that high fidelities can be achieved without applying any special procedures, such as fine-tuning of the shuttling path.
\end{abstract}

\maketitle
%TC:endignore

\textit{Introduction.}---Spin qubits in Si/SiGe heterostructures have a number of potential advantages, including their small size and long coherence times \cite{Loss:1998p120, Zwanenburg:2013p961, Burkard:2023p025003}.
High-fidelity single- and two-qubit gates have been demonstrated \cite{Yoneda:2018p102, Xue:2022p343, Noiri:2022p338, Mills:2022p5130}, and device yield and uniformity continue to improve as device designs and process nodes are optimized \cite{Ha:2022p1443, Neyens:2024p80}.

\begin{figure}[t] 
	\includegraphics[width=8cm]{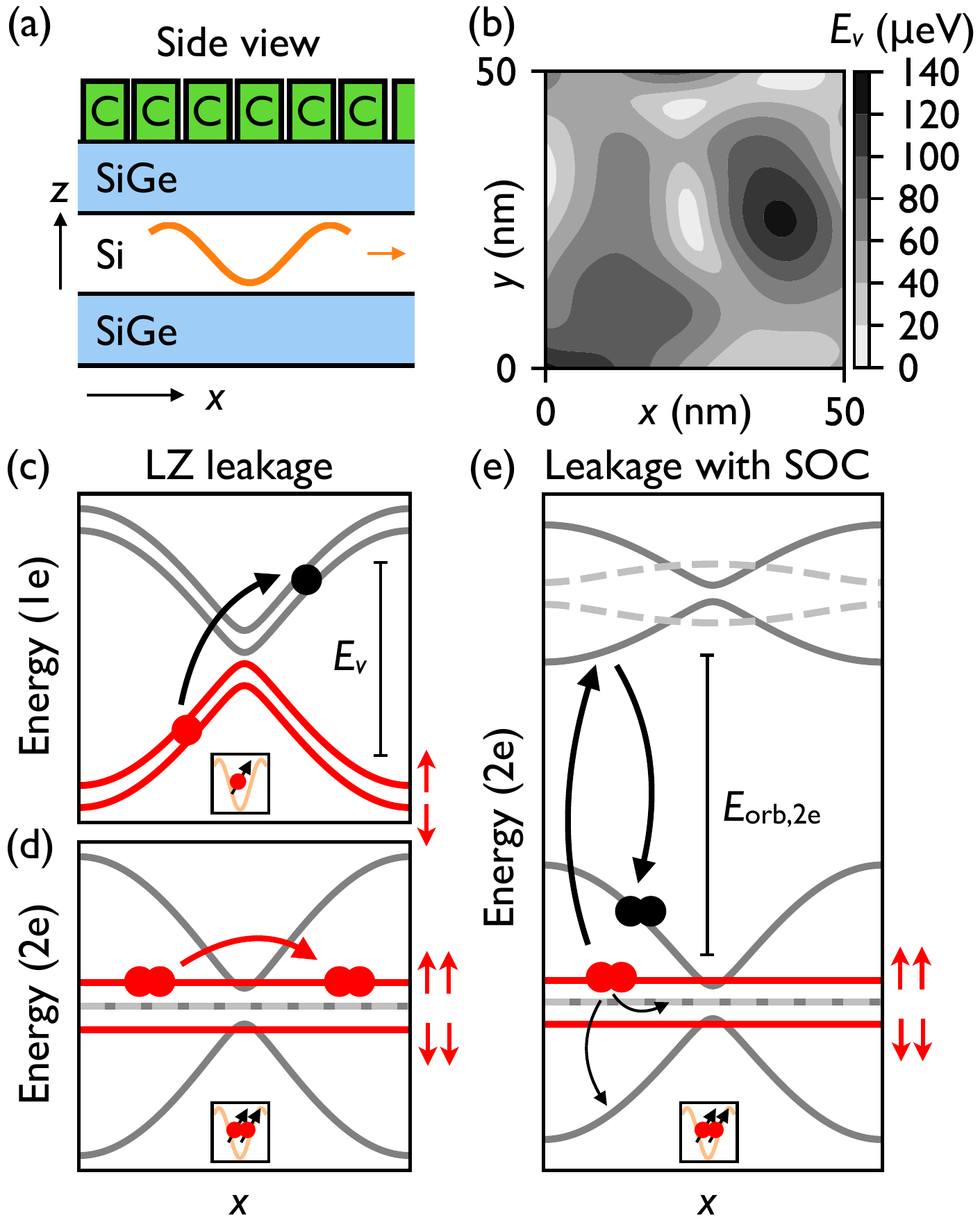}
	\centering
	\caption{Encoding schemes for high-fidelity conveyor-mode shuttling in silicon.
    (a) Cross-sectional schematic of the shuttling device, including clavier gates (C) fabricated above a Si/SiGe quantum well.
    Time-varying electrostatic potentials create pockets that can carry electrons along the shuttling channel. 
    (b) Realistic variations of the valley energy splitting $E_v$ across a $50\times50$~\SI{}{\nano\meter\squared} region of heterostructure, generated as in \cite{Losert:2024p040322, Woods:2025arXiv}. 
    (c) Energy levels versus position $x$ of a single-electron dot.  
    When the dot travels through a region with small valley splitting $E_v$, Landau-Zener (LZ) processes can excite the electron into an excited valley state.
    (d) Energies versus position $x$ of a two-electron dot with the same single-electron valley splitting as in (c). 
    Valley singlets do not experience anticrossings, and are largely immune to leakage. 
    In this work, we explore the spin-polarized spin-triplet encoding, shown in red.
    (e) Leakage is the dominant error mechanism when shuttling valley singlets, including a weak first-order process enabled by spin-orbit coupling (thin arrows), and stronger second-order processes involving the virtual occupation of excited orbital states (thick arrows).}
	\label{fig:intro}
\end{figure}

Spin qubits in quantum well heterostructures have natural nearest-neighbor connectivity arising from exchange interactions.
Much progress has also been made in engineering longer-scale connectivity between qubits in semiconductor heterostructures over the scale of microns \cite{Trifunovic:2012p011006, Friesen:2007p230503, Braakman:2013p432, Fujita:2017p22, Serina:2017p245422, Tosi:2017p450, Landig:2018p179, Samkharadze:2018p1123, Mills:2019p113501, Warren:2019p161303, Sigillito:2019p110, Yoneda:2021p4114, Qiao:2021p017701, Holman:2021p137, Jadot:2021p570, Noiri:2022p5740, Boter:2022p024053, Wang:2023p2208557, Zwerver:2023p030303, White:2024arXiv}. 
One particularly promising technology is the conveyor-mode shuttling architecture \cite{Taylor:2005p177, Seidler:2022p100, Langrock:2023p020305, Ermoneit:2023preprint, Struck:2024p1325, Xue:2024p2296, Volmer:2024p61, Kunne:2024p4977, DeSmet:2024arXiv, Mokeev:2024arXiv, Jeon:2025p195302, Oda:2026p020802, David:2024arXiv}.
In this scheme, oscillating sinusoidal potentials are applied to clavier gates interleaved across a shuttling channel, creating potential pockets that can drive an electron across a device, as illustrated schematically in Fig.~\ref{fig:intro}(a).

A critical challenge for Si/SiGe spin qubits, and conveyor-mode shuttling in particular, is the presence of low-lying valley states.
The band structure of a strained Si quantum well has two conduction-band minima, with nearly degenerate valley states.
The energy gap between these states, called the valley splitting $E_v$, varies across a given device, yielding a remarkably wide range of values between 1.5-\SI{300}{\micro\electronvolt} \cite{Zajac:2015p223507, Hollmann:2020p034068, Borselli:2011p063109, Shi:2011p233108, Scarlino:2017p165429, Mi:2017p176803, Ferdous:2018p26, Mi:2018p161404, Neyens:2018p243107, Borjans:2019p044063, Oh:2021p125122, Chen:2021p044033, Volmer:2024p61, Marcks:2025p11381, Volmer:2025arXiv}.
Theoretical work has demonstrated that random-alloy disorder is the dominant source of these fluctuations \cite{Wuetz:2022p7730, Losert:2023p125405, Lima:2023p025004, Pena:2024p33, Marcks:2025p11381}.
A typical disorder-dominated valley energy landscape is shown in Fig.~\ref{fig:intro}(b).
Importantly, this disorder creates regions of vanishing valley splitting, scattered across a device \cite{Losert:2024p040322, Woods:2025arXiv}.
If a shuttled dot passes through such a region, a Landau-Zener-like excitation can cause it to leak outside of the computational subspace, as illustrated in Fig.~\ref{fig:intro}(c) \cite{Losert:2024p040322, Lima:2025p235439, Volmer:2025arXiv}.
Such low-$E_v$ regions are a primary challenge for conveyor-mode shuttling in Si, motivating much theoretical work in the field \cite{Losert:2024p040322, Lima:2025p235439, David:2024arXiv, Oda:2024arXiv, Nemeth:2025arXiv}.

Multi-electron qubit encodings have been proposed to protect shuttlers from specific noise sources. 
For example, the two-electron $S$-$T_0$ encoding (spin-singlet and unpolarized-triplet) was proposed to reduce magnetic noise sensitivity \cite{Langrock:2023p020305, Zhang:2025p205301}, while three-electron spin-$\frac{1}{2}$ states were proposed to screen spurious electric fields \cite{Langrock:2023p020305}.
In this work, we propose an alternative scheme to protect qubits from valley-state excitations. 
As illustrated in Fig.~\ref{fig:intro}(d), the encoded basis is formed of two electrons in a single dot, having the same ground-orbital and valley-singlet states.
(Here, we assume that valley splitting is smaller than orbital splitting). 
The logical states are then defined within the polarized-spin-triplet subspace.
A simple mapping exists between conventional Loss-DiVincenzo qubits and this two-electron encoding, as described in the Supplemental Materials (SM).
The success of the scheme relies on the fact that direct transitions from the logical states to other states in the low-energy manifold requires a spin flip [indicated by thin arrows in Fig.~\ref{fig:intro}(e)], which is a very weak process in Si.
Second-order processes involving virtually excited orbital states [indicated by thick arrows in Fig.~\ref{fig:intro}(e)] are therefore dominant, although they are suppressed by large orbital excitation energies \cite{Langrock:2023p020305}.
In contrast with single-electron shuttling, we show here that the proposed scheme actually \emph{benefits} from low valley splittings.
Moreover, in certain regimes, high-fidelity shutting can be achieved \textit{without} fine-tuning the shuttling path---also in contrast with single-electron shuttling.

\textit{Physical model.}---We begin by considering the single-electron problem, ignoring the spin for now, before moving to the two-electron case.
We use the effective-mass approximation (EMA) to describe the valley physics, in which single-electron valley states are formed from linear combinations of $k$-states $|z_\pm\rangle$, representing the Si conduction-band minima at points $\pm k_0 \hat{\bf z}$ in the Brillouin zone, where $k_0 = 0.82 \times 2\pi / a_0$ and $a_0 = 0.543$~\SI{}{\nano\meter} is the lattice constant of the silicon diamond cubic lattice \cite{Friesen:2007p115318}.
In the $|z_\pm\rangle$ basis, the single-electron valley Hamiltonian is given by $H_v = \epsilon_0 \tau_0 + \text{Re}[\Delta_0] \tau_x - \text{Im}[\Delta_0] \tau_y$, where $\tau_j$ are Pauli operators acting on the valley basis $\{|z_\pm\rangle\}$.
To first order, the inter-valley coupling $\Delta_0$ and intra-valley coupling $\epsilon_0$ are given by
\begin{equation} \label{eq:delta_main}
\begin{split}
    \Delta_0 & \coloneqq \langle z_- | U_\text{qw} | z_+ \rangle = \int d\mathbf{r} \; e^{-2 i k_0 z} \psi_0^2 U_\text{qw} , \\
    \epsilon_0 & \coloneqq \langle z_{\pm} | U_\text{qw} | z_\pm \rangle = \int d\mathbf{r} \; \psi_0^2 U_\text{qw} ,
\end{split}
\end{equation}
where $\psi_0$ is the ground-state orbital envelope function (often, assumed to be the ground-state of a harmonic confinement potential) and the quantum-well confinement potential $U_\text{qw}$ is determined by the local concentration of Ge atoms across the Si/SiGe heterostructure.
(We note that $U_\text{qw}$ includes contributions from random-alloy disorder, nonzero interface widths, and atomic steps in the interface.)
The valley splitting is then given by $E_v = 2|\Delta_0|$ \cite{Friesen:2007p115318}.
For quantum wells with realistic interface thicknesses of greater than or equal to 3 atomic monolayers, $\Delta_0$ is mainly determined by alloy disorder, and is thus randomized \cite{Losert:2023p125405}.
In this limit, $\Delta_0$ is well-described as an isotropic, complex, Gaussian random variable with zero mean and variance $\sigma_{\Delta_0}^2$, for which the real and imaginary components are independent.
In general, the value of $\sigma_{\Delta_0}$ depends on the quantum-well profile and confinement strength \cite{Losert:2023p125405}; however, the heterostructure details are not critical to our model, and it is sufficient to take $\sigma_{\Delta_0}$ as the only disorder parameter in our theory. 
The two-electron valley-coupling variance $\sigma_{\Delta_s}$ is thus a derived quantity proportional to $\sigma_{\Delta_0}$, which we define in the SM.

The one-electron Hamiltonian $H_v$ can be transformed from the static $|z_\pm\rangle$ basis to the basis of ground and excited valley states $|v_{g(e)}\rangle$, through the rotation
\begin{equation} \label{eq:U_val_1e}
    U_v = \frac{1}{\sqrt{2}} 
    \begin{pmatrix}
        -e^{-i \phi_0} & 1 \\ e^{-i \phi_0} & 1
    \end{pmatrix} ,
\end{equation}
where $\phi_0 = \text{Arg}[\Delta_0]$.
Since $\phi_0$ (and $E_v$) vary across the quantum well, this is not a static transformation during shuttling; the effective time-dependent single-electron Hamiltonian therefore acquires a dynamical correction $-i \hbar U_v \dot U_v^\dag$, resulting in  
$\tilde H_v = -(E_v/2) \tilde \tau_z  - (\hbar \dot \phi_0 / 2) \tilde \tau_x + (\hbar \dot \phi_0 / 2)  \tilde \tau_0$.
(Here and throughout this work, we use tildes to indicate quantities in the basis of instantaneous valley eigenstates.)
Valley phase fluctuations therefore couple the ground and excited valley states through the $\tilde \tau_x$ term, yielding Landau-Zener-like excitations. 
In contrast with conventional Landau-Zener excitations, however, the $\tilde \tau_z$ and $\tilde \tau_x$ terms in $\tilde H_v$ are both time-dependent here. 
Excitations therefore occur near a valley splitting minimum when the level velocity is large, $\hbar\dot E_v/2\pi(E_v^\text{min})^2\gtrsim 1$, as illustrated in Fig.~\ref{fig:intro}(c), \emph{or} when the phase velocity is large \cite{Lima:2025p235439}, $\hbar\dot\phi/2\pi E_v^\text{min}\gtrsim 1$.

We now consider the two-electron system.
There are two main differences with the case of single electrons.
First, electron-electron interactions modify the envelope functions and reduce the orbital splitting compared to the single-electron case \cite{Corrigan:2021p127701, Ercan:2021p235302, Ercan:2022p247701, Abadillo-Uriel:2021p195305, Yannouleas:2022p21LT01, Yannouleas:2022p205302, Yannouleas:2022p195306, Ercan:2023arXiv}, although we assume the lateral confinement is strong enough that excited orbital states are well-separated and may be treated perturbatively (see SM).
Second, we account for Fermion parity, as follows.
Our system has three relevant degrees of freedom (orbital, valley, and spin) and six states in the lowest-energy manifold, as illustrated in Fig.~\ref{fig:intro}(e).
Three of these are valley-singlet/spin-triplet states: $|0\rangle = |S^\text{val} T_-^\text{spin} \rangle$, $|1 \rangle = |S^\text{val} T_+^\text{spin} \rangle$, and $|2 \rangle = |S^\text{val} T_0^\text{spin} \rangle$, where we define $|S^\text{val}\rangle = (\ket{z_+ z_-} - \ket{z_- z_+ })/\sqrt{2}$, $|T_-^\text{spin}\rangle = \ket{ \downarrow \downarrow }$, $|T_+^\text{spin}\rangle = \ket{\uparrow \uparrow }$, and $|T_0^\text{spin} \rangle = (\ket{\uparrow \downarrow} + \ket{\downarrow\uparrow }) / \sqrt{2}$. 
The remaining three are valley-triplet/spin-singlet states: $\ket{3} = |T_-^\text{val} S^\text{spin} \rangle$, $\ket{4} = |T_+^\text{val} S^\text{spin} \rangle$, and $\ket{5} = |T_0^\text{val} S^\text{spin} \rangle$, where $|T_-^\text{val} \rangle = \ket{z_- z_-}$, $|T_+^\text{val} \rangle = \ket{z_+ z_+}$, $|T_0^\text{val} \rangle = (\ket{z_+ z_-} + \ket{z_- z_+ })/\sqrt{2}$, and $|S^\text{spin} \rangle = (\ket{\uparrow \downarrow} - \ket{\downarrow\uparrow }) / \sqrt{2}$.
Note that the two-electron envelope functions $\psi_s$, which include the effects of electron-electron interactions, are identical for all six of these states, allowing us to drop their orbital indices for brevity.
Before introducing couplings to higher orbital states and spin-orbit couplings, the spin-valley Hamiltonian is therefore given by $H_\text{sv}^{2e} = \epsilon_s - E_z |0\rangle \langle 0| + E_z |1 \rangle \langle 1| + \{ \sqrt{2} (\Delta_s |3 \rangle \langle 5| + \Delta_s^*|4 \rangle \langle 5|) + h.c. \}$, where \emph{h.c.}\ denotes the Hermitian conjugate, and $\Delta_s$ and $\epsilon_s$ are defined analogously to Eq.~(\ref{eq:delta_main}), with the single-electron envelope $\psi_0$ replaced by the two-electron envelope $\psi_s$ (see SM).
We also define the two-electron quantum-well potential as $U_\text{qw}^{2e} = U_\text{qw}^{(1)} \otimes I^{(2)} + I^{(1)} \otimes U_\text{qw}^{(2)}$, where the superscripts label the electrons.

We include spin-orbit coupling (SOC) in our model following Refs.~\cite{Woods:2023p035418, Woods:2024arXiv}, taking into account both inter-valley and intra-valley Rashba and Dresselhaus terms (see SM).
Although SOC is quite weak for electrons in Si/SiGe quantum dots, it plays an important role in spin shuttling by creating a leakage channel.
SOC generally depends on both the shuttling direction and the magnetic field orientation; here, we assume an in-plane magnetic field $\mathbf{B} = B \hat x$, parallel to the shutting axis, as explained below, where $B = 20$~\SI{}{\milli\tesla}.
Additionally, we ignore inter-valley $g$-factor variations, which are equal and opposite for the ground and excited valley states~\cite{Woods:2024arXiv}, therefore canceling out for our logical basis states.
For the other states in the low-energy manifold, the $g$-factor fluctuations are small compared to the dominant energy splittings, allowing us to ignore them.

Similar to our approach for single electrons, we now diagonalize $H_\text{sv}^{2e}$ by applying a rotation between the valley basis states $\{ |S^\text{val}\rangle, |T_-^\text{val}\rangle, |T_+^\text{val}\rangle, |T_0^\text{val} \rangle \}$, given by
\begin{equation} \label{eq:urot_valley}
U_v^{2e} = 
\begin{pmatrix}
-e^{-i \phi_s} & 0 & 0 & 0\\
0 & \frac{1}{2} e^{-2 i \phi_s} & \frac{1}{2} & -\frac{1}{\sqrt{2}} e^{-i \phi_s} \\
0 &  \frac{1}{2} e^{-2 i \phi_s} & \frac{1}{2} & \frac{1}{\sqrt{2}} e^{-i \phi_s} \\
0 & -\frac{1}{\sqrt{2}} e^{-2 i \phi_s} & \frac{1}{\sqrt{2}} & 0
\end{pmatrix} ,
\end{equation}
where $\phi_s  = \text{Arg}[\Delta_s]$.
A key observation is that $U_v^{2e}$ leaves $|S^\text{val}\rangle$ decoupled from other valley states.
It is this property that ensures the stability of valley singlets against Landau-Zener leakage.
As in the single-electron case, $U_v^{2e}$ acquires a time dependence during shuttling, due to the spatial variation of $\phi_s$, which induces dynamical corrections in the transformed Hamiltonian.

\begin{figure*}[t] 
	\includegraphics[width=16cm]{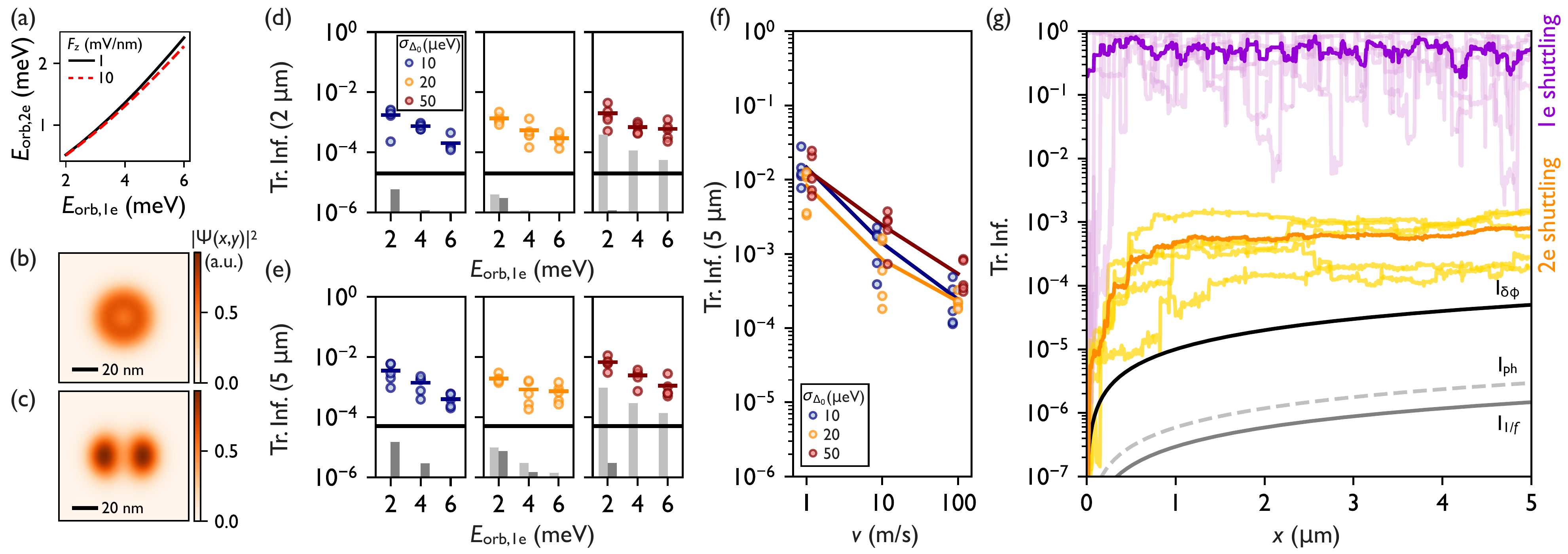}
	\centering
	\caption{
    Infidelities obtained when shuttling spin triplets (Tr. Inf.) for different disorder and confinement strengths. 
    (a) Two-electron orbital energies as a function of the confinement strength $E_{\text{orb},1e}$, computed using FCI methods.
    Here, we assume vertical fields of $F_z = 1$ and \SI{10}{\milli\volt\per\nano\meter} and do not include alloy disorder. 
    (b) Ground-state and (c) first-excited-state orbital densities, for $E_{\text{orb},1e} = 2$~\SI{}{\milli\electronvolt} and $F_z = 1$~\SI{}{\milli\volt\per\nano\meter}, again computed with no disorder. 
    (d) Trace infidelities computed at a shuttling distance of \SI{2}{\micro\meter}, for a variety of confinement and disorder strengths ($E_{\text{orb},1e}$ and $\sigma_{\Delta_0}$, respectively), assuming a shuttling velocity of $v = 10$~\SI{}{\meter\per\second}. 
    Open circles show simulation results for five randomly generated disorder landscapes, at each parameter setting; horizontal tick marks show the corresponding average values.
    We also plot infidelity estimates obtained for dephasing (black lines), electron-phonon coupling (light-gray bars), and $1/f$ noise (dark-gray bars).
    (e) Same as (d), for a shuttling distance of \SI{5}{\micro\meter}. 
    (f) Shuttling infidelities at a final distance of \SI{5}{\micro\meter}, for $E_{\text{orb},1e} = 4$~\SI{}{\milli\electronvolt}, obtained at three different shuttling velocities. 
    (g) The five trace infidelities shown in (d) and (e), for the case $E_{\text{orb},1e} = 4$~\SI{}{\milli\electronvolt} and $\sigma_{\Delta_0} = 20$~\SI{}{\micro\electronvolt}, are replotted here as a function of distance (light yellow); the corresponding average is plotted in dark yellow. 
    For comparison, we also plot five one-electron trace infidelities for the same shuttling parameters (light purple), with their corresponding average (dark purple). 
    Infidelity estimates are shown for dephasing (black line), electron-phonon coupling (dashed line), and $1/f$ noise (gray line).}
	\label{fig:fidelity}
\end{figure*}

In this way, we obtain an effective first-order, two-electron Hamiltonian of the form
\begin{multline} 
\label{eq:ham_first_order}
    \tilde H_\text{sv}^{2e} = (2 \epsilon_s + \hbar \dot \phi_s) I_{6\times 6} \\
    +\begin{pmatrix}
        -E_z  & 0 & \sqrt{2}m_\text{so} & 0 & 0& 0 \\
        0 & E_z & \sqrt{2}m_\text{so}^* & 0 & 0 & 0 \\
        \sqrt{2}m_\text{so}^* & \sqrt{2}m_\text{so} & 0 & 0 & 0 & 0  \\
        0 & 0 & 0 & -E_{v,2e} & 0 & -\frac{\hbar \dot \phi_s}{\sqrt{2}} \\
        0 & 0 & 0 & 0 & E_{v,2e}& -\frac{\hbar \dot \phi_s}{\sqrt{2}} \\
        0 & 0 & 0 & -\frac{\hbar \dot \phi_s}{\sqrt{2}} & -\frac{\hbar \dot \phi_s}{\sqrt{2}} & 0
    \end{pmatrix} ,
\end{multline}
where we define $E_{v,2e} = 2|\Delta_s|$ and $m_\text{so} = i m_t v \alpha / \hbar$, as well as the shuttling speed $v$, the Rashba spin-orbit coupling parameter $\alpha = 2$~\SI{}{\micro\electronvolt\nano\meter}, and the transverse effective mass $m_t = 0.19 m_e$, where $m_e$ is the bare electron mass.
[We also include a Dresselhaus SO term $\beta = 12$~\SI{}{\micro\electronvolt\nano\meter}, which does not appear in Eq.~(\ref{eq:ham_first_order}), due to our choice of field direction, but plays an important role in the second-order leakage effects, described below.]
The first two rows of $\tilde H_\text{sv}^{2e}$ represent the computational states, $|0_L\rangle = |\tilde S^\text{val}T_-^\text{spin}\rangle$ and $|1_L\rangle = |\tilde S^\text{val} T_+^\text{spin}\rangle$, whose valley components are expressed in terms of instantaneous eigenstates: $|\tilde S^\text{val} \rangle = (|v_e v_g \rangle - |v_g v_e \rangle) / \sqrt{2}$.
The remaining rows represent the leakage states, $|l_2\rangle = |\tilde S^\text{val} T_0^\text{spin}\rangle$, $|l_3\rangle = |\tilde T_-^\text{val} S^\text{spin} \rangle$, $|l_4\rangle = |\tilde T_+^\text{val} S^\text{spin} \rangle$, and $|l_5\rangle = |\tilde T_0^\text{val}  S^\text{spin} \rangle$.

\begin{comment}
Electrons in Si/SiGe quantum dots are also known to experience weak spin-orbit coupling, which nonetheless plays an important role in spin shuttling.
We use the spin-orbit coupling model of Woods et al.~\cite{Woods:2023p035418, Woods:2024unpublished}, including both inter-valley and intra-valley Rashba and Dresselhaus terms.
For a single particle in the $z_\pm$ valley basis, $  H_\text{so} = \alpha \tau_0 (k_x \sigma_y - k_y \sigma_x) + (\beta \tau_- + \beta^* \tau_+) (k_x \sigma_x - k_y \sigma_y)$, where $\sigma_i$ are Pauli matrices in spin space, $\tau_i$ are Pauli matrices in the $z_\pm$ valley basis, $\tau_\pm = (\tau_x \pm i \tau_y)/2$ are valley raising and lowering operators, and $\alpha$ and $\beta = |\beta| e^{i\phi_\beta}$ are the Rashba and Dresselhaus spin-orbit coupling parameters.
We take $\alpha = 2$~\SI{}{\micro\electronvolt\nano\meter} and $|\beta| = 12$~\SI{}{\micro\electronvolt\nano\meter} \cite{Woods:2023p035418}, and we set $\phi_\beta = 0$ for simplicity.
The equivalent two-electron Hamiltonian is $H_\text{so}^{2e} = H_\text{so}^{(1)} \otimes I^{(2)} + I^{(1)} \otimes H_\text{so}^{(2)}$.
Transforming to a frame co-moving with the quantum dot, we replace $k_j \rightarrow k_j + m_t v_j / \hbar$, where $v_j$ is the dot velocity along axis $j \in \{ x,  y\}$.
After applying the valley rotation of Eq.~(\ref{eq:urot_valley}), we obtain an effective spin-orbit Hamiltonian in our low-energy subspace.
\end{comment}

Some comments are in order for $\tilde H_\text{sv}^{2e}$.
As in the single-electron case, the dynamical terms (proportional to $\dot\phi_s$) cause a coupling between the valley states.
Crucially however, only the $\tilde T^\text{val}$ states are affected.
The disordered valley landscape, which generates the $\dot \phi_s$ terms, therefore does not directly couple the logical states $|0_L\rangle$ and $|1_L\rangle$.
On the other hand, SOC couples the logical states to $|l_2\rangle$, creating a weak leakage channel.
Dresselhaus SOC could also potentially couple the logical states to $|l_3\rangle$, $|l_4\rangle$, and $|l_5\rangle$.
Our choice of magnetic field orientation suppresses these couplings here, although they are generally weak and would not significantly affect our results (see SM).
In contrast, second-order virtual couplings to excited orbital states present a stronger leakage channel, as we now show.

\textit{Second-order effects.}---The six basis states considered so far are formed with both electrons in their low-energy $s$ orbitals.
We now consider couplings to higher-lying $p_x$ and $p_y$ orbitals, separated from the $s$ orbital by the energy scale $E_{\text{orb},2e}$ [see Fig.~\ref{fig:intro}(e)].
Since $E_{\text{orb},2e}$ is large compared to other shuttling energies, any excitation above the $s$ manifold must be virtual, resulting in effective second-order couplings.
The inter-orbital coupling terms $\langle s | H | p \rangle$ are of two types:
(1) SOC processes, which flip spins (and can also involve valleys), and (2) valley-orbit coupling (VOC) processes, which arise from alloy disorder.
Accounting for both $p_x$ and $p_y$ states, there are 32 excited orbital levels that satisfy Fermion parity constraints.
Removing these states from the Hamiltonian via a Schrieffer-Wolff transformation gives second-order couplings of the form \cite{WinklerBook}
\begin{equation} \label{eq:ham_second_order}
    \tilde H'_{i j} = \frac{1}{2} \sum_k \tilde H_{ik} \tilde H_{kj} \left( \frac{1}{E_i - E_k} + \frac{1}{E_j - E_k} \right) ,
\end{equation}
where $i$ and $j$ label states within the $s$ manifold, $k$ labels states in the $p_x$ and $p_y$ manifolds, $\tilde H_{ik}$ are matrix elements between the manifolds (evaluated in the instantaneous valley basis), and $E_i$ are the instantaneous eigenvalues of the first-order Hamiltonian.
There are a total of $6 \times 32 = 192$ matrix elements, so we leave the details of the calculation to the SM.
However, we comment on the scaling behavior of the leakage terms here.
First, we note that all second-order terms are proportional to $1/E_{\text{orb},2e}$, so leakage is suppressed for more strongly-confined systems, with larger $E_{\text{orb},2e}$.
Next, we note that SOC matrix elements are proportional to $\beta k_{sp}$, where $\beta$ is a (Rashba or Dresselhaus) SO parameter and $k_{sp}=\langle s | k | p \rangle$. 
Intra-valley and inter-valley matrix elements are proportional to integrals of the form $\int \! d\mathbf{r}\, \psi_s \psi_p U_\text{qw}$ and $\int\! d\mathbf{r} \, e^{-2 i k_0 z} \psi_s \psi_p U_\text{qw}$, respectively, which are both of order $\sigma_{\Delta_0}$.
The main second-order processes responsible for leakage involve one SOC transition and one VOC transition, resulting in a single spin flip; the resulting leakage terms in $\tilde H'$ therefore scale as $\beta k_{sp} \sigma_{\Delta_0} / E_{\text{orb},2e}$.
On the whole, we see that our scheme benefits from silicon's low SOC, and can be further suppressed by reducing valley-scattering effects associated with alloy disorder.

For the simulations described below, we have used a dot confinement potential $U_\text{conf} = \frac{1}{2} m_t \omega^2 [(x-x_0)^2 + (y-y_0)^2] + e F_z z + U_\text{qw}^\text{vc}(z)$, where $(x_0, y_0)$ is the dot center,  $E_{\text{orb},1e} = \hbar \omega$ is the characteristic single-electron confinement energy, $F_z$ is the vertical electric field, and $U_\text{qw}^\text{vc}$ is the quantum well potential in the virtual crystal approximation (i.e., without random-alloy disorder). 
We assume a \SI{10}{\nano\meter}-wide quantum well, with interface widths of \SI{0.8}{\nano\meter}, as consistent with state-of-the-art fabrication techniques \cite{Wuetz:2022p7730, Esposti:2024p32} (see SM).

In two-electron systems, the orbital excitation energy $E_{\text{orb},2e}$ is suppressed (relative to the single-electron orbital energy $E_{\text{orb},1e}$) by electron-electron interactions \cite{Ercan:2021p235302, Ercan:2022p247701}.
To account for this, we solve the interacting two-electron problem using the effective-mass approximation and the full configuration interaction (FCI) module of the MaSQE software package \cite{Anderson:2022p065123}.
The orbital energies at two different vertical fields, $F_z = 1$~\SI{}{\milli\volt\per\nano\meter} and \SI{10}{\milli\volt\per\nano\meter}, are shown in Fig.~\ref{fig:fidelity}(a).
In Figs.~\ref{fig:fidelity}(b) and \ref{fig:fidelity}(c), we also plot the ground and first-excited orbital wavefunctions corresponding to $E_{\text{orb},1e} = 2$~\SI{}{\milli\electronvolt} and $F_z = 1$~\SI{}{\milli\volt\per\nano\meter}, highlighting their distinct $s$ and $p$ characters.
For the model chosen here, we note that $p_x$ and $p_y$ are energetically degenerate.
In the dynamical simulations described below, we use the $E_{\text{orb},2e}$ value obtained by averaging FCI results for $F_z = 1$~\SI{}{\milli\volt\per\nano\meter} and \SI{10}{\milli\volt\per\nano\meter}.

%We consider the confinement potential $U_\text{conf} = \frac{1}{2} m_t \omega^2 [(x-x_0)^2 + (y-y_0)^2] + e F_z z + U_\text{qw}^\text{vc}(z)$, where $(x_0, y_0)$ is the dot center,  $E_{\text{orb},1e} = \hbar \omega$ is the characteristic single-electron confinement energy, $F_z$ is the vertical electric field, and $U_\text{qw}^\text{vc}$ is the quantum well potential in the virtual crystal approximation (i.e., without random-alloy disorder). 
%We assume a \SI{10}{\nano\meter}-wide quantum well, with interface widths of \SI{0.8}{\nano\meter}, as consistent with state-of-the-art fabrication techniques \cite{Wuetz:2022p7730, Esposti:2024p32} (see SM).

%We also consider two different vertical fields, $F_z = 1$ and \SI{10}{\milli\volt\per\nano\meter}, obtaining the results shown in Fig.~\ref{fig:fidelity}(a).

\textit{Shuttling simulations.}---To evaluate the performance of our spin-triplet shuttling scheme, and to benchmark it against single-electron shuttling, we perform simulations of a \SI{5}{\micro\meter} shuttler.
As noted above, in the disordered regime, once the quantum dot confinement potential is specified, all statistical properties related to valley coupling are captured in a single parameter, $\sigma_{\Delta_0}$, which in turn determines the parameters $\Delta_0$, $\Delta_s$, and $\sigma_{\Delta_s}$.
These quantities are used to randomly generate spatially varying matrix elements in our simulations, as detailed in the SM.
For each disorder landscape, we numerically solve the Schr\"odinger equation for the time-evolution propagator, $i \hbar \dot U = H U$. 
In the two-electron case, we have $H = \tilde H_\text{sv}^{2e} + \tilde H'$, where $\tilde H_\text{sv}^{2e}$ is given in Eq.~(\ref{eq:ham_first_order}) and $\tilde H'$ is given in Eq.~(\ref{eq:ham_second_order}).
Thus, we obtain a large set of simulation results for different instantiations of random-alloy disorder.
For each solution, we evaluate the shuttling performance by computing the trace infidelity \cite{Palao:2002p188301} $I = 1 - \left[ \frac{1}{2} | \Tr P U P U_\text{id}^\dag  | \right]^2$ 
where $P = |0_L\rangle \langle 0_L| + |1_L \rangle \langle 1_L|$ projects the system onto the logical subspace, and $U_\text{id}$ represents the ``ideal'' evolution, in the absence of disorder and leakage, obtained by removing all SO and VO couplings.
These simulations are implemented in the Julia programming language \cite{Julia:2017, JuliaDiffEq:2017}.

The results of these two-electron shuttling simulations are shown in Figs.~\ref{fig:fidelity}(d) and \ref{fig:fidelity}(e), where we plot the trace infidelities for shutting distances of 2 or \SI{5}{\micro\meter}, respectively, with a fixed shuttling velocity of $v_x = 10$~\SI{}{\meter\per\second}.
Results are shown for three different disorder levels, $\sigma_{\Delta_0}= \{10, 20, 50\}$~\SI{}{\micro\electronvolt} (color coded), and three confinement strengths, $E_{\text{orb},1e}=\{2,4,6\}$~\SI{}{\milli\electronvolt}.
[We find that disorder levels as low as $\sigma_{\Delta_0}=10$~\SI{}{\micro\electronvolt} are achievable in realistic devices by tuning $F_z$; see SM.]
Each group of simulations is repeated for five different disorder landscapes, all corresponding to the same $\sigma_{\Delta_0}$ value.
(Individual simulation results are shown as colored circles, with averages shown as horizontal colored ticks.)
As expected, we find that infidelities decrease (fidelities improve), on average, as $E_{\text{orb},1e}$ increases or as $\sigma_{\Delta_0}$ decreases.
For reference, we also include estimates of the infidelity due to dephasing, $I_{\delta \phi}$ (long horizontal black lines), calculated as in Ref.~\cite{Losert:2024p040322}.
Since our qubit encoding is composed of excited valley-singlet states, we also estimate the infidelity caused by $T_1$-type decay, due to either electron-phonon coupling (light gray bars) or $1/f$ noise (dark gray bars). 
The details of these calculations are described in the SM; however, we emphasize that leakage errors are always found to be the dominant error source, here.

We also explore how infidelities scale with the shuttling velocity. 
In Fig.~\ref{fig:fidelity}(f), we show averaged simulation results for velocities $v_x = 1$, 10, and \SI{100}{\meter\per\second}, with $E_{\text{orb},1e} = 4$~\SI{}{\milli\electronvolt} and three disorder levels, $\sigma_{\Delta_0}= \{10, 20, 50\}$~\SI{}{\micro\electronvolt} (also color coded). 
Here, we find that infidelities decrease with higher shuttling speeds, reflecting behavior known as \emph{motional narrowing}, in which the higher speeds cause the noise effects to be ``averaged out'' and suppressed \cite{Langrock:2023p020305}.
For the highest velocity considered in these simulations ($v_x=100$~\SI{}{\meter\per\second}), we find that shuttling infidelities less than $10^{-3}$ are consistently attainable in the strongly confined, small-$E_v$ regime.

As a benchmark, we also simulate single-electron shuttling.
In this case, the logical basis is defined as the ground-valley spin states.
The single-electron Hamiltonian is simply given by $H^{1e} = \tilde H_v + E_z \sigma_z / 2$, which does not include SOC or couplings to higher orbital levels, since they do not contribute to the Landau-Zener-like excitations that strongly dominate the single-electron shuttling infidelity.
The single-electron trace infidelity is defined similarly to the two-electron case, by projecting onto the appropriate logical basis.

In Fig.~\ref{fig:fidelity}(g), we compare the trace-infidelity results for one- and two-electron shuttling simulations using the \emph{same} alloy-disorder landscapes obtained with the parameters $\sigma_{\Delta_0} = 20$~\SI{}{\micro\electronvolt}, $E_{\text{orb},1e} = 4$~\SI{}{\milli\electronvolt}, and $v_x=10$~\SI{}{\meter\per\second}.
Here, our previous two-electron results are plotted in yellow, as a function the shuttling distance $x$, with the average of the five simulations plotted in gold.
Five analogous single-electron results are plotted in light purple, with their average plotted in dark purple.
The two-electron coding scheme is seen to yield fidelity improvements of 2-3 orders of magnitude over the whole shuttling range, compared to the conventional single-electron scheme, substantiating our proposal.
%Of course, fidelities can be improved by allowing for transverse motion or other techniques \cite{Losert:2024p040322}; however, none of these techniques are used in the present simulations.
For reference in Fig.~\ref{fig:fidelity}(g), we also include estimates of infidelities cause by dephasing $I_{\delta \phi}$ (black line), phonon-induced decay $I_\text{ph}$ (light-gray dashed line), and $1/f$ noise-induced decay (solid gray line) (see SM for details).
Typically, these errors are one or more orders of magnitude smaller than leakage errors, justifying our choice to ignore them in our simulations.

\textit{Summary.}---In this work, we have proposed a  two-electron coding scheme for spin shuttling in realistic Si devices with disordered valley landscapes, based on spin-polarized triplets.
We show that shuttling in this system is largely immune to Landau-Zener valley excitations.
Instead, leakage is dominated by relatively weak second-order virtual excitations to excited orbital states, which can be suppressed by engineering dots with larger orbital splittings or smaller valley-orbit couplings.
In contrast with standard encodings, our approach therefore benefits from \emph{smaller} valley splittings.
For realistic shuttling parameters, we find that infidelities less than $10^{-3}$ should be achievable, with no fine-tuning of the shuttling path.
This is in contrast with other recent shuttling proposals that require both high valley splittings and tunability to achieve high fidelities.
We hope this work encourages future studies of unconventional qubit encodings, which may have desirable properties yet to be discovered.\\

%\textit{Data Availability.}---
%Source data and code presented in this work is available in a Zenodo repository \cite{Losert:TripletZenodo}. \\

\textit{Acknowledgements.}---
We thank Chris Anderson and Mark Gyure for help setting up and implementing the MaSQE FCI software package.
This research was performed in part while M.L. held an NRC
Research Associateship award at the National Institute of
Standards and Technology (NIST).
This research was sponsored in part by the Army Research Office (ARO) under Awards No.\ W911NF-17-1-0274, W911NF-22-1-0090, and W911NF-23-1-0110. 
The views, conclusions, and recommendations contained in this document are those of the authors and are not necessarily endorsed nor should they be interpreted as representing the official policies, either expressed or implied, of the Army Research Office or the U.S. Government. 
The U.S. Government is authorized to reproduce and distribute reprints for Government purposes, notwithstanding any copyright noted herein. 
Any mention of commercial products is for informational purposes only; it does not imply recommendation or endorsement by the National Institute of Standards and Technology.
Manuscript preparation was also supported in part through the National Science Foundation QLCI-HQAN (Award No. 2016136). Any opinions, findings, and conclusions or recommendations expressed in this material are those of the author(s) and do not necessarily reflect the views of the National Science Foundation.

\clearpage

\onecolumngrid

% For Supplementary Sections (e.g., S1, S2)
\setcounter{section}{0}
\renewcommand{\thesection}{S\arabic{section}}

% For Supplementary Figures (e.g., S1, S2)
\setcounter{figure}{0}
\renewcommand{\thefigure}{S\arabic{figure}}
\renewcommand{\figurename}{Figure} % Optional: Keeps it as "Figure S1"

% For Supplementary Equations (e.g., S1, S2)
\setcounter{equation}{0}
\renewcommand{\theequation}{S\arabic{equation}}

% For Supplementary Tables
\setcounter{table}{0}
\renewcommand{\thetable}{S\arabic{table}}

\section*{Supplementary Material}

\section{FCI simulations} \label{app:fci}
We utilize two forms of full-configuration-interaction simulations in this work, both performed within the MaSQE software package, based on the simulations of Ref.~\cite{Anderson:2022p065123}.
It is well-known that few-electron quantum dot wavefunctions are strongly affected by Coulomb interactions \cite{Corrigan:2021p127701, Ercan:2021p235302, Ercan:2022p247701, Abadillo-Uriel:2021p195305, Yannouleas:2022p21LT01, Yannouleas:2022p205302, Yannouleas:2022p195306, Ercan:2023arXiv}.
Hence, many-body techniques like FCI are needed to obtain accurate solutions.
The first technique, which we dub ``effective mass (EM)'' FCI, does not account for the valley physics.
We use this technique to obtain envelope functions, since the resulting wavefunctions do not contain fast valley oscillations, and to determine the orbital splittings in a two-electron quantum dot, since the level structure of the EM FCI model does not contain additional valley energies.
The second technique, which we dub ``tight-binding (TB)'' FCI, does account for valley physics, and employs the two-band tight-binding model of Boykin et al.~\cite{Boykin:2004p115, Boykin:2008p245320} as the kinetic energy operator.
We use this technique to directly probe the valley splitting in disordered heterostructures.
In this section, we outline the procedure used in both of these simulations.

\subsection{EM FCI simulations}

For the EM FCI simulations, we start with the Hamiltonian
\begin{equation} \label{eq:ham_2e_fci_supp}
    H_\text{EM}^{2e} = \sum_{i=1}^2 H_\text{EM}^{1e}(\mathbf{r}_i) + \frac{e^2}{4 \pi \epsilon_0 \epsilon_r} \frac{1}{|\mathbf{r}_1 - \mathbf{r}_2|}
\end{equation}
where $H_\text{EM}^{1e} = T_\text{EM} + U_\text{es} + U_\text{qw}^\text{vc}$.
$T_\text{EM}$ is a discretized kinetic energy operator that takes into account the longitudinal and transverse effective masses in Si, and the remaining terms are confinement potentials, including electrostatic confinement $U_\text{es} = \frac{1}{2} m_t \omega^2 [(x-x_0)^2 + (y-y_0)^2] + e F_z z$, with both lateral parabolic confinement and a vertical electric field, and the quantum well potential $U_\text{qw}^\text{vc}$.
The single-particle orbital spacing $E_{\text{orb},1e} = \hbar \omega$.
The quantum well potential is computed in the virtual crystal approximation, without alloy disorder:
\begin{equation}\label{eq:U_qw_vc}
    U_\text{qw}^\text{vc} = \Delta E_c\frac{\bar X_l - 1}{ X_s - 1 }
\end{equation}
where $\bar X_l$ is the expected Si concentration at layer $l$, averaged over the whole device, $X_s = 0.7$ is the Si concentration in the SiGe barrier (substrate), and $\Delta E_c = 150.6$~\SI{}{\milli\electronvolt} is the conduction band offset between the relaxed $\text{Si}_{0.7} \text{Ge}_{0.3}$ quantum well barrier and the strained Si quantum well \cite{Schaffler:1997p1515, Wuetz:2022p7730}.
We model the Si concentration profile as a sigmoid function,
\begin{equation} \label{eq:avg_si_conc_supp}
    \bar X_l = 1 + \frac{X_s - 1}{1 + \exp [(z - z_t)/\tau]} + \frac{X_s - 1}{1 + \exp [(z_b - z)/\tau]}
\end{equation}
where the quantum well top and bottom interface positions are given by $z_t$ and $z_b$, the well width $W = |z_b - z_t| = 10$~\SI{}{\nano\meter} and the interface widths $\lambda_\text{int} = 4 \tau = 6$~monolayer or \SI{0.8}{\nano\meter}, consistent with state-of-the-art fabrication techniques \cite{Wuetz:2022p7730, Esposti:2024p32}.
These FCI simulations ignore the valley degree of freedom entirely, as well as valley-orbit effects due to steps in the quantum well interface.
%For our regime of interest, these are both suitable approximations to obtain the first-order orbital energies, as explained below.
In order to simulate the system, we discretize our crystal lattice into rectangular cells of size $(\Delta x, \Delta y, \Delta z) = (2 a_0, 2 a_0, a_0/4)$. 
The quantum well and electrostatic confinement potentials are added as onsite parameters.

\subsection{TB FCI simulations} 
For the TB FCI simulations, we adopt the same form of the two-electron FCI Hamiltonian as Eq.~(\ref{eq:ham_2e_fci_supp}), where $H_\text{TB}^{1e} = T_\text{TB} + U_\text{es} + U_\text{qw}$. 
In the TB simulations, the kinetic operator $T_\text{TB}$ is given by the two-band tight-binding model of Boykin et al.~in the $\hat z$ direction, which reproduces the position and effective mass of valley states along the $\pm z$ axis of the Brillouin zone \cite{Boykin:2004p115, Boykin:2008p245320}.
Furthermore, since these simulations aim to capture valley physics that inherently depends on atomic disorder, we do not employ the virtual crystal approximation in defining $U_\text{qw}$. 
Instead, we adopt a model that accounts for alloy disorder.
Again discretizing the system into cells of size $(\Delta x, \Delta y, \Delta z) = (2 a_0, 2 a_0, a_0/4)$, we model the quantum well potential as
\begin{equation}
U_\text{qw}(j,k,l) = \Delta E_c \frac{1 - X_{jkl}}{1 - X_s}.
\end{equation}
where $X_{jkl}$ is the true Si concentration in the cell with indices $(j,k,l)$.
Since each cell in our lattice contains 8 atoms in the diamond cubic crystal lattice of Si/SiGe, we sample $X_{jkl}$ from a binomial distribution according to 
\begin{equation} \label{eq:binom}
X_{jkl} \sim \text{Binom}(\bar X_{jkl}, n_c) / n_c
\end{equation}
where $\text{Binom}(p, n)$ is a binomial distribution of $n$ trials with $p$ probability of success, $\bar X_{jkl}$ is the expected average Si concentration in the cell, and $n_c$ is the number of Si/Ge atoms per cell.
For our cell size, $n_c = 8$.
For systems without an interface step, we set $\bar X_{jkl} = \bar X_l$, where $\bar X_l$ is defined in Eq.~(\ref{eq:avg_si_conc_supp}).
For systems with a step, we define
\begin{equation} \label{eq:avg_si_conc_step_supp}
    \bar X_{jkl} = \bar X_l \Theta \left( x \leq x_\text{step} \right) + \bar X_{l+1} \Theta \left(x > x_\text{step} \right)
\end{equation}
where once again $\bar X_l$ is given in Eq.~(\ref{eq:avg_si_conc_supp}) and $\Theta(\cdot)$ is the Heaviside step function.

\section{Detailed description of the model Hamiltonian} \label{app:ham}

\begin{figure}[t] 
\centering
\begin{minipage}{0.30\linewidth}
  \caption{\label{fig:energies}
    Energy level structure of two-electron quantum dots. 
    (a) Schematic illustration of the energy levels in a doubly-occupied quantum dot, including the logical basis (red) and low-lying leakage states. The 8 single-electron excited orbital states are also illustrated: here, each two-electron excited orbital state is formed from one of the eight single-electron excited states and one low-energy state. 
    (b) The computed energies of the six low-lying instantaneous eigenstates across a heterostructure, for a randomly generated disorder landscape, including the two logical states $|S^\text{val} T_\pm^\text{spin} \rangle$ (red), and the remaining low-energy leakage states. We have ignored dynamical terms $\propto \dot \phi$ to compute these levels, and we have subtracted off a uniform constant offset $2\epsilon_s$ from each level. For this plot, we assume the parameters $\sigma_{\Delta_0} = 20$~\SI{}{\micro\electronvolt}, $E_{\mathrm{orb},1e} = 4$~\SI{}{\milli\electronvolt}, and $B = 20$~\SI{}{\milli\tesla}.}
\end{minipage}\hfill
\begin{minipage}{0.65\linewidth}
  \includegraphics[width=\linewidth]{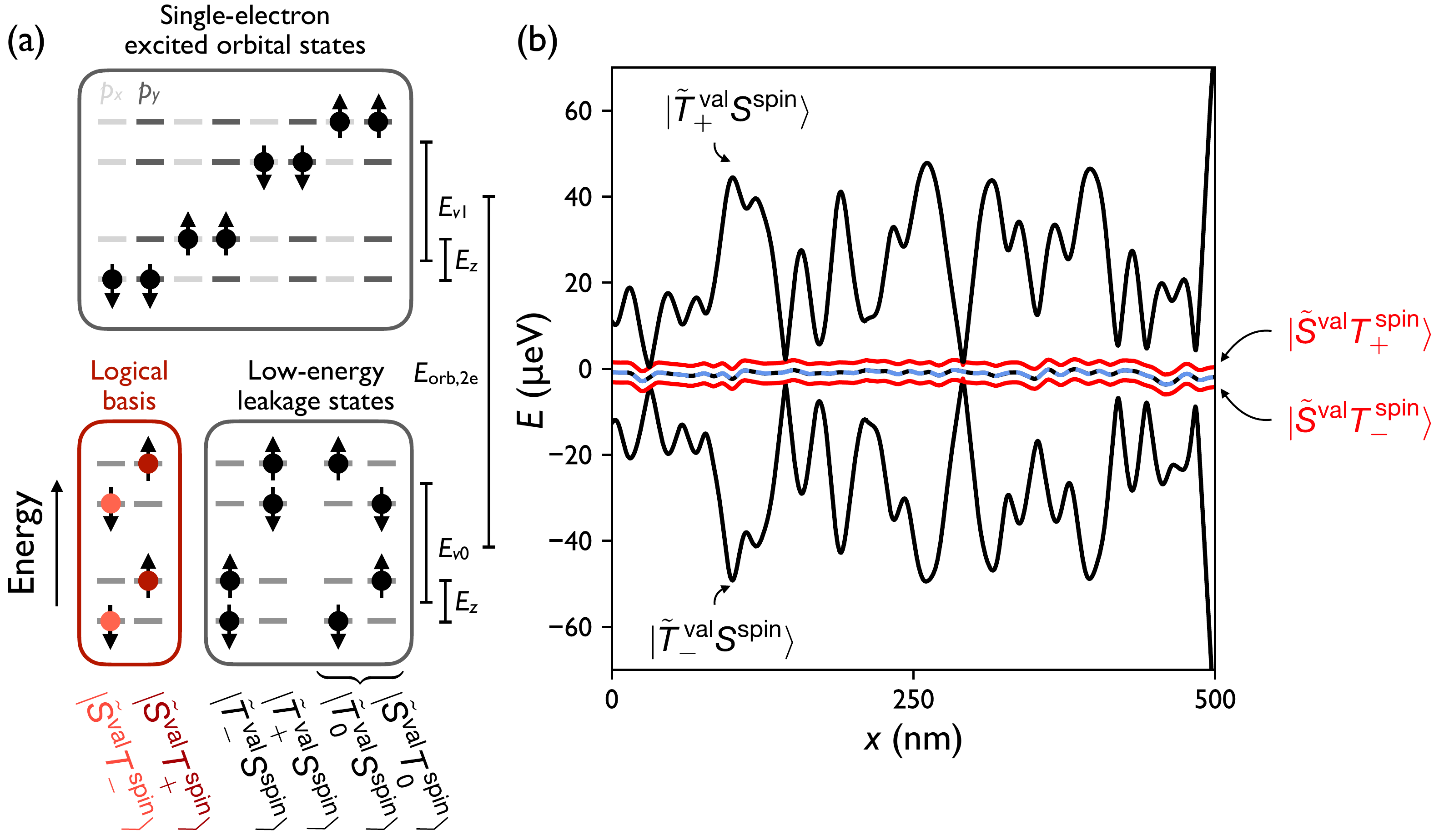}
\end{minipage}
\end{figure}

In this section, we provide details of the model Hamiltonian used in our shuttling simulations.
There are several important components of this Hamiltonian.
The primary dynamics are contained within the six-level low-energy subspace detailed in the main text.
In Sec.~\ref{app:low_e_so}, we also detail the spin-orbit coupling terms within this subspace, as reported in Eq.~(\ref{eq:ham_first_order}) of the main text.
There are two mechanisms that can couple low-energy states to excited orbital states: (1) valley-orbit coupling induced by alloy disorder in the quantum well, and (2) spin-orbit coupling, which we consider in Secs.~\ref{app:excited_vo} and \ref{app:excited_so} respectively.
These second-order couplings lead to leakage out of the qubit subspace, so it is crucial to account for these states correctly.
Then, in Sec.~\ref{app:excited_subspace}, we describe our model of the excited orbital subspace, comprised of states with a single orbital excitation.
Finally, in Sec.~\ref{app:schrieffer_wolff}, we describe the second-order Schrieffer-Wolff transform used to obtain the low-energy dynamics of the system.

\subsection{Spin-orbit coupling within the low-energy subspace} \label{app:low_e_so}

To compute the low-energy spin-orbit Hamiltonian, we need to compute matrix elements of the form $\langle L | H_\text{so}^{2e} | L' \rangle$, where $|L\rangle$ and $|L'\rangle$ both come from the six-level low-energy subspace.
The two-electron spin-orbit interaction decomposes into single-electron components, 
\begin{equation} \label{eq:2e_ham}
    H_\text{so}^{2e} = H_\text{so}^{(1)} \otimes I^{(2)} + I^{(1)} \otimes H_\text{so}^{(2)}
\end{equation}
where we model the spin-orbit coupling based on the model of Woods et al.~\cite{Woods:2024arXiv},
\begin{equation} \label{eq:ham_so_app}
      H_\text{so} = \alpha \tau_0 (k_x \sigma_y - k_y \sigma_x) + (\beta \tau_- + \beta^* \tau_+) (k_x \sigma_x - k_y \sigma_y).
\end{equation}
Here, $\tau_j$ are Pauli operators in valley space $\{z_-, z_+ \}$,  $\alpha$ is the Rashba spin-orbit parameter, and $\beta$ is the Dresselhaus spin-orbit parameter.
We compute two-electron matrix elements by expressing the basis states as product states and applying Eq.~(\ref{eq:2e_ham}).
For example,
\begin{equation}
\begin{split}
    \langle T_-^\text{spin}  S^\text{val} |  H_\text{so}^{2e} | S^\text{spin}  T_-^\text{val} \rangle &= \frac{1}{2} \left[ \langle \downarrow z_+; \downarrow z_- | - \langle \downarrow z_-; \downarrow z_+ |  \right]  H_\text{so}^{2e} \left[ |\uparrow z_-; \downarrow z_-\rangle - |\downarrow z_-; \uparrow z_-\rangle\right] 
    \\
    & = \frac{1}{2} \left[ \langle \downarrow z_+; \downarrow z_- |  H_\text{so}^{2e} | \uparrow z_-; \downarrow z_-\rangle  + \langle \downarrow z_-; \downarrow z_+ |  H_\text{so}^{2e} | \downarrow z_-; \uparrow z_- \rangle \right] 
    \\
    & = \frac{1}{2} \left[ \langle \downarrow z_+ |  H_\text{so} | \uparrow z_-\rangle \langle \downarrow z_- | \downarrow z_- \rangle + \langle \downarrow z_- | \downarrow z_- \rangle \langle \downarrow z_+ |  H_\text{so} | \uparrow z_- \rangle \right] 
    \\
    & = \langle \downarrow z_+ |  H_\text{so} | \uparrow z_- \rangle .
\end{split} 
\end{equation}
Transforming to a frame co-moving with the quantum dot, we replace $k_j \rightarrow k_j + m_t v_j / \hbar$.
Terms proportional to $k$ vanish within the low-energy subspace, leaving just the velocity-dependent terms.
In the static valley basis, we obtain the spin-orbit Hamiltonian
\begin{equation}
    H_\text{so} = 
    \begin{pmatrix}
        0 & 0 & \sqrt{2} \alpha R^{\downarrow \uparrow} & \beta^* D^{\downarrow \uparrow} & -\beta D^{\downarrow \uparrow} & 0 \\
        0 & 0 & \sqrt{2} \alpha R^{\uparrow \downarrow} & -\beta^* D^{\uparrow \downarrow} & \beta D^{\uparrow \downarrow} & 0 \\
        \sqrt{2} \alpha {R^{\downarrow \uparrow}}^* & \sqrt{2} \alpha {R^{\uparrow \downarrow}}^* & 0 & 0 & 0 & 0 \\
        \beta {D^{\downarrow \uparrow}}^* & -\beta {D^{\uparrow\downarrow}}^* & 0 & 0 & 0 & 0 \\
        -\beta^* {D^{\downarrow \uparrow}}^* & \beta^* {D^{\uparrow \downarrow}}^* & 0 & 0 & 0 & 0 \\
        0 & 0 & 0 & 0 & 0 & 0
    \end{pmatrix} ,
\end{equation}
where the Rashba and Dresselhaus components are given by
\begin{equation}
\begin{split}
    R^{i j} = \frac{m_t}{\hbar} \left(v_x \sigma_y^{ij} - v_y k_x^{ij} \right) , \\
    D^{i j} = \frac{m_t}{\hbar} \left( v_x \sigma_x^{ij} - v_y \sigma_y^{ij} \right) ,
\end{split}
\end{equation}
and the spin matrix elements are given by $\sigma_\nu^{ij} = \langle i | \sigma_\nu | j \rangle$, for $i, j \in \{ \uparrow, \downarrow \}$.
For a magnetic field $\mathbf{B} = B ( \sin \theta_B \cos \phi_B, \sin \theta_B \sin \phi_B, \cos \theta_B)$, the spin eigenstates are given by
\begin{equation}
\begin{split}
    |\downarrow \rangle &= - e^{-i \phi_B / 2} \sin(\theta_B / 2) | \uparrow_z \rangle + e^{i \phi_B / 2} \cos(\theta_B/2) | \downarrow_z \rangle \\
    |\uparrow \rangle &= e^{-i \phi_B/ 2} \cos(\theta_B / 2) |\uparrow_z \rangle + e^{i \phi_B / 2} \sin(\theta_B / 2) |\downarrow_z \rangle ,
\end{split}
\end{equation}
where $|\uparrow_z \rangle$ and $|\downarrow_z\rangle$ are the eigenstates of $\sigma_z$.
For an in-plane magnetic field ($\theta_B = \pi/2$), the spin matrix elements are given by
\begin{equation} \label{eq:spin_mat_el}
    \begin{split}
        \sigma_x^{\uparrow\uparrow} &= -\sigma_x^{\downarrow\downarrow} = \cos \phi_B = 1\\ 
        \sigma_x^{\uparrow\downarrow} & = \left(\sigma_x^{\downarrow\uparrow} \right)^* = i \sin \phi_B = 0\\
        \sigma_y^{\uparrow\uparrow} &= - \sigma_y^{\downarrow\downarrow} = \sin \phi_B = 0\\
        \sigma_y^{\uparrow\downarrow} &= \left(\sigma_y^{\downarrow\uparrow}  \right)^* = -i \cos \phi_B = -i .
    \end{split}
\end{equation}
In this work, we set $B$ to be along $\hat x$, so that $\phi_B = 0$, resulting in the values reported on the right-hand-side of Eq.~(\ref{eq:spin_mat_el}).
In our simulations in the main text, we set $v_x = v$ and $v_y = 0$.
If we apply the valley rotation operator $U_v$ of Eq.~(\ref{eq:urot_valley}) of the main text, we obtain the spin-orbit terms reported in the main text.
For nonzero $v_x$ and $v_y$ (and $B$ along $\hat x$) we obtain
\begin{equation}
    \tilde H_\text{so} = 
    \begin{pmatrix}
        0 & 0 & \frac{i \sqrt{2} \alpha m_t v_x}{\hbar}  & -\frac{|\beta| m_t v_y}{\hbar}\sin( \phi - \phi_\beta) & -\frac{|\beta| m_t v_y}{\hbar}\sin( \phi - \phi_\beta) & -\frac{i \sqrt{2}|\beta| m_t v_y}{\hbar}\cos( \phi - \phi_\beta) \\
        \cdot & 0 & -\frac{i \sqrt{2} \alpha m_t v_x}{\hbar}  & -\frac{|\beta| m_t v_y}{\hbar}\sin( \phi - \phi_\beta) & -\frac{|\beta| m_t v_y}{\hbar}\sin( \phi - \phi_\beta) & -\frac{i \sqrt{2}|\beta| m_t v_y}{\hbar}\cos( \phi - \phi_\beta) \\
        \cdot & \cdot & 0 & 0 & 0 & 0 \\
        \cdot & \cdot & \cdot & 0 & 0 & 0 \\
        \cdot & \cdot & \cdot & \cdot & 0 & 0 \\
        \cdot & \cdot & \cdot & \cdot & \cdot & 0
    \end{pmatrix}.
\end{equation}
The remaining terms are found by Hermitian conjugation.

\subsection{Disorder-induced valley-orbit coupling} \label{app:excited_vo}

Here, we examine the coupling of the 6-level low-energy subspace to excited orbital states through disorder-induced valley-orbit coupling.
To do so, we first need a model of these orbital states.
First, we consider the ground-orbital manifold. 
In these states, to lowest order, both electrons in the dot share the same envelope function, which we label $\psi_s$.
Since both electrons share the same envelope, we can straightforwardly determine $\psi_s$ using EM FCI simulations.
We label the resulting two-electron orbital wavefunction
\begin{equation}
    \ket{T_-^\mathrm{orb}} = \ket{\psi_s^{(1)} \psi_s^{(2)}}.
\end{equation}

Next, we consider excited states with a single orbital excitation.
These states are composed of two single-electron wavefunctions, where one is $s$-like and one is $p$-like.
We note that the $s$-like single-electron envelope function in an excited orbital state is, in general, not equal to $\psi_s$, since the different charge distribution of the $p$-state modifies the Coulomb potential.
For simplicity, we assume that these excited orbital states are composed of one simple harmonic oscillator ground state $\psi_0$ and one SHO $p$-state, $\psi_{p_\mu}$, where $\mu \in \{ x, y \}$:
\begin{equation} \label{eq:SHO}
    \begin{split}
        \psi_{0}(x,y) & = \left( \frac{m_t \omega}{\pi \hbar} \right)^{1/4} \exp \left[-\frac{x^2 + y^2}{2 a_\text{dot}^2} \right] \\
        \psi_{p_\mu}(x,y) & = \left( \frac{m_t \omega}{\pi \hbar} \right)^{1/4} \left[ \frac{\sqrt{2} \mu}{a_\text{dot}} \right] \exp \left[-\frac{x^2 + y^2}{2 a_\text{dot}^2} \right].
    \end{split}
\end{equation}
Such a model is, of course, approximate, but it allows us to derive statistical descriptions of all couplings used in this work, as we describe below.
The resulting two-electron orbital wavefunctions have either singlet or triplet character:
\begin{equation}
\begin{split}
    |S^\mathrm{orb}_\mu \rangle &= \frac{1}{\sqrt{2}} \left( \ket{\psi_{p_\mu}^{(1)} \psi_0^{(2)}} - \ket{\psi_0^{(1)} \psi_{p_\mu}^{(2)} } \right)
    \\
    |T_{0,\mu}^\mathrm{orb} \rangle &= \frac{1}{\sqrt{2}} \left( \ket{\psi_{p_\mu}^{(1)} \psi_0^{(2)}} + \ket{\psi_0^{(1)} \psi_{p_\mu}^{(2)} } \right).
\end{split}
\end{equation}
We assume the $z$ components of $\psi_s$, $\psi_0$, and $\psi_{p_\mu}$ are identical.
With $x$- and $y$-excitations combined with spin and valley degrees of freedom, we have 32 excited states in total, illustrated schematically in Fig.~\ref{fig:energies}(a).

Armed with these envelope functions, we can compute matrix elements of the quantum well disorder potential.
This disorder is spin-independent, so all such matrix elements preserve spin.
As described above, all two-particle matrix elements reduce to single-electron integrals.
We compute these couplings in the static valley basis, with valley states $\{z_-, z_+\}$.
First, we compute matrix elements coupling our low-energy subpsace with higher orbital states.
There are two types of matrix elements: valley-preserving terms, and valley-flipping terms.
The non-zero valley-preserving terms are given by
\begin{equation} 
\langle T_-^\text{orb}| U_\text{qw} | T_{0,\mu}^\text{orb} \rangle = 
\sqrt{2} \left( O_{sp_\mu} \epsilon_{ss'} + O_{s0} \epsilon_{sp_\mu} \right) ,
\end{equation}
where $\mu \in \{x, y\}$, depending on whether the excited orbital state is of $x$ or $y$ type, the single-particle integrals $O$ and $\epsilon$ are defined below, and we have dropped valley and spin labels for clarity.
The valley-flipping terms are given by
\begin{equation} 
\begin{split}
\langle T_-^\text{orb}  S^\text{val} | U_\text{qw} | S_\mu^\text{orb}  T_-^\text{val} \rangle &= \Delta^*_{sp_\mu} O_{s0}  - \Delta^*_{s0} O_{sp_\mu} 
\\
\langle T_-^\text{orb}  S^\text{val} | U_\text{qw} | S_\mu^\text{orb}  T_+^\text{val} \rangle &= \Delta_{s0} O_{sp_\mu}  - \Delta_{sp} O_{s0} 
\\
\langle T_-^\text{orb}  S^\text{val} | U_\text{qw} | S_\mu^\text{orb}  T_0^\text{val} \rangle &= 0
\\
\langle T_-^\text{orb}   T_-^\text{val} | U_\text{qw} | T_{0,\mu}^\text{orb}  T_0^\text{val} \rangle &= \Delta_{sp_\mu} O_{s0}  + \Delta_{s0} O_{sp_\mu} 
\\
\langle T_-^\text{orb}  T_+^\text{val} | U_\text{qw} | T_0^\text{orb}  T_{0,\mu}^\text{val} \rangle &=  \Delta^*_{sp_\mu} O_{s0}  + \Delta^*_{s0} O_{sp_\mu}
\\
\langle T_-^\text{orb}  T_0^\text{val} | U_\text{qw} | T_{0,\mu}^\text{orb}  T_+^\text{val} \rangle &= \Delta_{sp_\mu} O_{s0}  + \Delta_{s0} O_{sp_\mu}
\\
\langle T_-^\text{orb}  T_0^\text{val} | U_\text{qw} | T_{0,\mu}^\text{orb}  T_-^\text{val} \rangle &= \Delta^*_{sp_\mu} O_{s0}  + \Delta^*_{s0} O_{sp_\mu}
\\
\langle T_-^\text{orb}  T_0^\text{val} | U_\text{qw} | S_\mu^\text{orb}  S^\text{val} \rangle &= 0.
\end{split}
\end{equation}
Alloy disorder also induces diagonal corrections to the orbital energies within both the ground and first-excited manifolds, given by
\begin{equation} \label{eq:diagonal_orb}
\begin{split}
\delta E_{s} &= \langle T_-^\mathrm{orb} | U_\mathrm{qw} | T_-^\mathrm{orb} \rangle = 2 \epsilon_s \\
\delta E_{p_\mu} &= \langle T_{0,\mu}^\mathrm{orb} | U_\mathrm{qw} | T_{0,\mu}^\mathrm{orb} \rangle = \langle S_{\mu}^\mathrm{orb} | U_\mathrm{qw} | S_{\mu}^\mathrm{orb} \rangle = \epsilon_{0} + \epsilon_{p_\mu} .
\end{split}
\end{equation}
We define the following disorder-induced single-particle integrals. 
First, the overlap terms:
\begin{equation} \label{eq:overlap_ints_sup}
\begin{split}
O_{s0} & \coloneqq \langle z_\pm, s  | z_\pm, 0\rangle = \int d\mathbf{r} \; \psi_s \psi_{0} \\
O_{sp_\mu} & \coloneqq \langle z_\pm, s  | z_\pm, p_\mu \rangle = \int d\mathbf{r} \; \psi_s \psi_{p_\mu}.
\end{split}
\end{equation}
Since we consider symmetric confinement potentials, we assume $O_{sp_\mu} = 0$.
For a given confinement potential, we compute $O_{s0}$ for $E_z = 1$ and \SI{10}{\milli\volt\per\nano\meter}, taking the average of the two. 
We find that $O_{s0} = \{0.93, 0.95, 0.97 \}$ for $E_{\text{orb},1e} = \{ 2, 4, 6\}$~\SI{}{\milli\electronvolt}.
Next, the valley-preserving single-particle integrals:
\begin{equation} \label{eq:valley_preserving_ints_supp}
\begin{split}
\epsilon_{s} & \coloneqq \langle z_\pm, s |U_\text{qw} | z_\pm, s \rangle = \int d\mathbf{r} \;  \psi_s^2 U_\text{qw} 
\\
\epsilon_{0} & \coloneqq \langle z_\pm, 0 | U_\text{qw} | z_\pm, 0 \rangle = \int d\mathbf{r} \;  \psi_{0}^2 U_\text{qw} 
\\
\epsilon_{p_\mu} & \coloneqq \langle z_\pm, p_\mu | U_\text{qw} | z_\pm, p_\mu \rangle = \int d\mathbf{r} \;  \psi_{p_\mu}^2 U_\text{qw} 
\\
\epsilon_{s0} & \coloneqq \langle z_\pm, s| U_\text{qw} | z_\pm, 0 \rangle = \int d\mathbf{r} \;  \psi_s \psi_{0}U_\text{qw} 
\\
\epsilon_{s p_\mu} & \coloneqq \langle z_\pm, s | U_\text{qw} | z_\pm, p_\mu \rangle = \int d\mathbf{r} \;  \psi_s \psi_{p_\mu}U_\text{qw} 
\\
\epsilon_{p_x p_y} & \coloneqq \langle z_\pm , p_x | U_\mathrm{qw} | z_\pm, p_y \rangle = \int d\mathbf{r} \; \psi_{p_x} \psi_{p_y} U_\mathrm{qw} .
\end{split}
\end{equation}
And finally, the valley-flipping single-particle integrals:
\begin{equation} \label{eq:valley_flipping_ints_supp}
\begin{split}
	\Delta_s & \coloneqq \langle z_-, s| U_\text{qw} | z_+, s \rangle = \int d\mathbf{r} \; e^{-2 i k_0 z} \psi_s^2 U_\text{qw} 
    \\
    \Delta_{0} & \coloneqq \langle z_- , 0 | U_\mathrm{qw} | z_+, 0 \rangle = \int d\mathbf{r} \; e^{-2 i k_0 z} \psi_{0}^2 U_\mathrm{qw}
    \\
    \Delta_{p_\mu} & \coloneqq \langle z_- , p_\mu | U_\mathrm{qw} | z_+, p_\mu \rangle = \int d\mathbf{r} \; e^{-2 i k_0 z} \psi_{p_\mu}^2 U_\mathrm{qw}
    \\
	\Delta_{s0}  & \coloneqq \langle z_-, s| U_\text{qw} | z_+, 0 \rangle = \int d\mathbf{r} \; e^{-2 i k_0 z} \psi_s \psi_{0}U_\text{qw} 
    \\
	\Delta_{sp_\mu} & \coloneqq  \langle z_-, s| U_\text{qw} | z_+, p_\mu \rangle = \int d\mathbf{r} \; e^{-2 i k_0 z} \psi_s \psi_{p_\mu}U_\text{qw} 
    \\
    \Delta_{p_x p_y} & \coloneqq \langle z_- , p_x | U_\mathrm{qw} | z_+, p_y \rangle = \int d\mathbf{r} \; e^{-2 i k_0 z} \psi_{p_x} \psi_{p_y} U_\mathrm{qw} .
\end{split}
\end{equation}
Thus, we can describe all valley-orbit coupling terms due to alloy disorder in terms of the integrals in Eqs.~(\ref{eq:overlap_ints_sup}), (\ref{eq:valley_preserving_ints_supp}), and (\ref{eq:valley_flipping_ints_supp}).

\subsection{Spin-orbit coupling to excited orbitals} \label{app:excited_so}

Now, we examine coupling to excited orbital states through the spin-orbit interaction. 
Once again, we start with the spin-orbit Hamiltonian given in Eq.~(\ref{eq:ham_so_app}).
This time, we examine coupling between different orbital manifolds. 
For these states, terms proportional to $v$ vanish, leaving just terms proportional to $k_{j}$. 
We use the definition
\begin{equation} \label{eq:k_mu}
    \langle s | k_\mu | p_\mu \rangle \coloneqq k_{sp_\mu} = -i \int d\mathbf{r} \; \psi_s \frac{\partial}{\partial \mu} \psi_p
\end{equation}
where $\mu \in \{ x, y \}$.
As with other parameters, we compute $k_{sp_\mu}$ numerically using $\psi_s$ and $\psi_p$ determined with vertical fields $F_z = 1$ and \SI{10}{\milli\volt\per\nano\meter}, and we use the average in our simulations.
Plugging into Eq.~(\ref{eq:ham_so_app}), we obtain the matrix elements coupling to excited orbital states.
Again, we work in the static valley basis, $\{ z_+, z_- \}$.
For example, spin-orbit matrix elements connecting $|T_-^\text{orb}  S^\text{val} T_-^\text{spin} \rangle$ to higher-energy orbital states are
\begin{equation} \label{eq:excited_orbital_SO_couplings}
\begin{split}
\langle T_-^\text{orb} S^\text{val} T_-^\text{spin} | H_\text{so} | S_\mu^\text{orb} S^\text{val} S^\text{spin}\rangle &= \frac{1}{2} O_{s0} \left( H_{--, \mu}^{\downarrow \uparrow} + H_{++, \mu}^{\downarrow \uparrow} \right)
\\
\langle T_-^\text{orb} S^\text{val} T_-^\text{spin} | H_\text{so} | S_\mu^\text{orb} T_-^\text{val} T_0^\text{spin} \rangle &= \frac{1}{\sqrt{2}} O_{s0} H_{+-,\mu}^{\downarrow \uparrow}
\\
\langle T_-^\text{orb} S^\text{val} T_-^\text{spin} | H_\text{so} | S_\mu^\text{orb} T_0^\text{val} T_0^\text{spin} \rangle &= \frac{1}{2} O_{s0} \left( H_{++,\mu}^{\downarrow \uparrow} - H_{--,\mu}^{\downarrow \uparrow}  \right) = 0
\\
\langle T_-^\text{orb} S^\text{val} T_-^\text{spin} | H_\text{so} | S_\mu^\text{orb}  T_+^\text{val} T_0^\text{spin} \rangle &= -\frac{1}{\sqrt{2}} O_{s0} H_{+-,\mu}^{\downarrow \uparrow}
\\
\langle T_-^\text{orb}  S^\text{val} T_-^\text{spin} | H_\text{so} | T_{0,\mu}^\text{orb}  S^\text{val} T_0^\text{spin} \rangle &= \frac{1}{2} O_{s0} \left( H_{--,\mu}^{\downarrow \uparrow} + H_{++,\mu}^{\downarrow \uparrow} \right)
\\
\langle T_-^\text{orb}  S^\text{val} T_-^\text{spin} | H_\text{so} | T_{0,\mu}^\text{orb}  T_-^\text{val} S^\text{spin} \rangle &= \frac{1}{\sqrt{2}} O_{s0} H_{+-,\mu}^{\downarrow \uparrow} 
\\
\langle T_-^\text{orb}  S^\text{val} T_-^\text{spin} | H_\text{so} | T_{0,\mu}^\text{orb}  T_0^\text{val} S^\text{spin} \rangle &= \frac{1}{2} O_{s0} \left( -H_{--,\mu}^{\downarrow \uparrow} + H_{++,\mu}^{\downarrow\uparrow} \right) = 0
\\
\langle T_-^\text{orb}  S^\text{val} T_-^\text{spin} | H_\text{so} | T_{0,\mu}^\text{orb}  T_+^\text{val} S^\text{spin} \rangle &= -\frac{1}{\sqrt{2}} O_{s0} H_{-+,\mu}^{\downarrow\uparrow}
\end{split}
\end{equation}
where once again $\mu \in \{ x, y\}$ labels the $p_x$ or $p_y$ character of the excited orbital state, and we note that the spin-orbit interaction does not couple $s$ and $0$ single-particle orbitals, since they are both rotationally symmetric about the origin.
The single-electron matrix elements $H_{z_i z_j,\mu}^{s_a s_b}$  are given by
\begin{equation}
\begin{split}
H_{++,x}^{s_a s_b} = H_{--,x}^{s_a s_b} &= -\alpha k_{sp_x} \sigma_y^{s_a s_b} \\
H_{++,y}^{s_a s_b} = H_{--,y}^{s_a s_b} &= \alpha k_{sp_y} \sigma_x^{s_a s_b} \\
H_{+-,x}^{s_a s_b} & = \beta e^{-i \phi_\beta} k_{sp_x} \sigma_x^{s_a s_b} \\
H_{+-,y}^{s_a s_b} & = -\beta e^{-i \phi_\beta} k_{sp_y} \sigma_y^{s_a s_b} \\
H_{-+,x}^{s_a s_b} & = \beta e^{i \phi_\beta} k_{sp_x} \sigma_x^{s_a s_b} \\
H_{-+,y}^{s_a s_b} & = -\beta e^{i \phi_\beta} k_{sp_y} \sigma_y^{s_a s_b} 
\end{split}
\end{equation}
where $s_a, s_b \in \{ \uparrow, \downarrow \}$, and $\sigma_{x(y)}^{s_a s_b}$ are given in Eq.~(\ref{eq:spin_mat_el}).
We repeat the above calculations for each pair of basis states, providing all the spin-orbit couplings between the ground and excited orbital manifolds.

\subsection{Excited orbital manifold}\label{app:excited_subspace}

Next, we discuss the matrix elements \textit{within} the excited orbital manifold, composed of orbital states with one excitation (i.e.~$T_{0,\mu}^\mathrm{orb}$ or $S_{\mu}^\mathrm{orb}$).
In this block, we ignore spin-orbit coupling, since terms within the excited orbital block will only impact the low-energy subspace at higher order.
Thus, we can compute all relevant matrix elements using the single-particle valley-orbit integrals defined in Sec.~\ref{app:excited_vo}.
First, the diagonal elements are given by $\delta E_{p_\mu} + E_{\mathrm{orb},2e}$, described in Eq.~(\ref{eq:diagonal_orb}).
The off-diagonal couplings \textit{within} the $p_x$ or $p_y$ manifolds are given by
\begin{equation}
\begin{split}
    \langle S^\mathrm{orb} T_-^\mathrm{val} | U_\mathrm{qw} | S^\mathrm{orb} T_+^\mathrm{val} \rangle & = 0
    \\
    \langle S^\mathrm{orb} T_-^\mathrm{val} | U_\mathrm{qw} | S^\mathrm{orb} T_0^\mathrm{val} \rangle & = \frac{1}{\sqrt{2}} \left( \Delta_0 + \Delta_{p_\mu} \right)
    \\
    \langle S^\mathrm{orb} T_-^\mathrm{val} | U_\mathrm{qw} | T_0^\mathrm{orb} S^\mathrm{val} \rangle & = \frac{1}{\sqrt{2}} \left( -\Delta_0 + \Delta_{p_\mu} \right)
    \\
    \langle S^\mathrm{orb} T_+^\mathrm{val} | U_\mathrm{qw} | S^\mathrm{orb} T_0^\mathrm{val} \rangle & = \frac{1}{\sqrt{2}} \left( \Delta_0^* + \Delta_{p_\mu}^* \right)
    \\
    \langle S^\mathrm{orb} T_+^\mathrm{val} | U_\mathrm{qw} | T_0^\mathrm{orb} S^\mathrm{val} \rangle & = \frac{1}{\sqrt{2}} \left( \Delta_0^* - \Delta_{p_\mu}^* \right)
    \\
     \langle S^\mathrm{orb} T_0^\mathrm{val} | U_\mathrm{qw} | T_0^\mathrm{orb} S^\mathrm{val} \rangle & = 0
    \\
    \langle T_0^\mathrm{orb} T_-^\mathrm{val} | U_\mathrm{qw} | T_0^\mathrm{orb} T_+^\mathrm{val} \rangle &= 0
    \\
    \langle T_0^\mathrm{orb} T_-^\mathrm{val} | U_\mathrm{qw} | T_0^\mathrm{orb} T_0^\mathrm{val} \rangle &= \frac{1}{\sqrt{2}} \left( \Delta_0 + \Delta_{p_\mu} \right)
    \\
    \langle T_0^\mathrm{orb} T_-^\mathrm{val}  | U_\mathrm{qw} | S^\mathrm{orb} S^\mathrm{val} \rangle &= \frac{1}{\sqrt{2}} \left( -\Delta_0 + \Delta_{p_\mu} \right)
    \\
    \langle T_0^\mathrm{orb} T_+^\mathrm{val}  | U_\mathrm{qw} | T_0^\mathrm{orb} T_0^\mathrm{val} \rangle &= \frac{1}{\sqrt{2}} \left( \Delta_0^* + \Delta_{p_\mu}^* \right)
    \\
    \langle T_0^\mathrm{orb} T_+^\mathrm{val}  | U_\mathrm{qw} | S^\mathrm{orb} S^\mathrm{val} \rangle &= \frac{1}{\sqrt{2}} \left(\Delta_0^* - \Delta_{p_\mu}^* \right)
    \\
    \langle T_0^\mathrm{orb} T_0^\mathrm{val}  | U_\mathrm{qw} | S^\mathrm{orb} S^\mathrm{val} \rangle &= 0,
\end{split}
\end{equation}
where we have suppressed the $\mu$ indices on the orbital states for clarity.
The couplings \textit{between} the $p_x$ block and the $p_y$ block are given by
\begin{equation}
\begin{split}
    \langle S^\mathrm{orb}_x T_-^\mathrm{val} | U_\mathrm{qw} | S^\mathrm{orb}_y T_-^\mathrm{val} \rangle & = \epsilon_{p_x p_y}
    \\
    \langle S^\mathrm{orb}_x T_-^\mathrm{val} | U_\mathrm{qw} | S^\mathrm{orb}_y T_+^\mathrm{val} \rangle & = 0
    \\
    \langle S^\mathrm{orb}_x T_-^\mathrm{val} | U_\mathrm{qw} | S^\mathrm{orb}_y T_0^\mathrm{val} \rangle & = \frac{1}{\sqrt{2}} \Delta_{p_x p_y}
    \\
    \langle S^\mathrm{orb}_x T_-^\mathrm{val} | U_\mathrm{qw} | T_{0,y}^\mathrm{orb} S^\mathrm{val} \rangle & = \frac{1}{\sqrt{2}} \Delta_{p_x p_y}
    \\
    \langle S^\mathrm{orb}_x T_+^\mathrm{val} | U_\mathrm{qw} | S_{y}^\mathrm{orb} T_+^\mathrm{val} \rangle & = \epsilon_{p_x p_y}
    \\
    \langle S^\mathrm{orb}_x T_+^\mathrm{val} | U_\mathrm{qw} | S_{y}^\mathrm{orb} T_0^\mathrm{val} \rangle & = \frac{1}{\sqrt{2}} \Delta_{p_x p_y}^*
    \\
    \langle S^\mathrm{orb}_x T_+^\mathrm{val} | U_\mathrm{qw} | T_{0,y}^\mathrm{orb} S^\mathrm{val} \rangle & =  -\frac{1}{\sqrt{2}} \Delta_{p_x p_y}^* 
    \\
    \langle S^\mathrm{orb}_x T_0^\mathrm{val} | U_\mathrm{qw} | S_{y}^\mathrm{orb} T_0^\mathrm{val} \rangle & = \epsilon_{p_x p_y}
    \\
    \langle S^\mathrm{orb}_x T_0^\mathrm{val} | U_\mathrm{qw} | T_{0,y}^\mathrm{orb} S^\mathrm{val} \rangle & = 0
    \\
    \langle T_{0,x}^\mathrm{orb} S^\mathrm{val} | U_\mathrm{qw} | T_{0,y}^\mathrm{orb} S^\mathrm{val} \rangle & = \epsilon_{p_x p_y}
    \\
    \langle T_{0,x}^\mathrm{orb} T_-^\mathrm{val} | U_\mathrm{qw} | T_{0,y}^\mathrm{orb} T_-^\mathrm{val} \rangle & = \epsilon_{p_x p_y}
    \\
    \langle T_{0,x}^\mathrm{orb} T_-^\mathrm{val} | U_\mathrm{qw} | T_{0,y}^\mathrm{orb} T_+^\mathrm{val} \rangle & = 0
    \\
    \langle T_{0,x}^\mathrm{orb} T_-^\mathrm{val} | U_\mathrm{qw} | T_{0,y}^\mathrm{orb} T_0^\mathrm{val} \rangle & = \frac{1}{\sqrt{2}}\Delta_{p_x p_y}
    \\
    \langle T_{0,x}^\mathrm{orb} T_-^\mathrm{val} | U_\mathrm{qw} | S_{y}^\mathrm{orb} S^\mathrm{val} \rangle & = \frac{1}{\sqrt{2}}\Delta_{p_x p_y}
    \\
    \langle T_{0,x}^\mathrm{orb} T_+^\mathrm{val} | U_\mathrm{qw} | T_{0,y}^\mathrm{orb} T_+^\mathrm{val} \rangle & = \epsilon_{p_x p_y}
    \\
    \langle T_{0,x}^\mathrm{orb} T_+^\mathrm{val} | U_\mathrm{qw} | T_{0,y}^\mathrm{orb} T_0^\mathrm{val} \rangle & = \frac{1}{\sqrt{2}} \Delta_{p_x p_y}^*
    \\
    \langle T_{0,x}^\mathrm{orb} T_+^\mathrm{val} | U_\mathrm{qw} | S_{y}^\mathrm{orb} S^\mathrm{val} \rangle & = -\frac{1}{\sqrt{2}} \Delta_{p_x p_y}^*
    \\
    \langle T_{0,x}^\mathrm{orb} T_0^\mathrm{val} | U_\mathrm{qw} | T_{0,y}^\mathrm{orb} T_0^\mathrm{val} \rangle & = \epsilon_{p_x p_y}
    \\
    \langle T_{0,x}^\mathrm{orb} T_0^\mathrm{val} | U_\mathrm{qw} | S_{y}^\mathrm{orb} S^\mathrm{val} \rangle & = 0
    \\
    \langle S_{x}^\mathrm{orb} S^\mathrm{val} | U_\mathrm{qw} | S_{y}^\mathrm{orb} S^\mathrm{val} \rangle & = \epsilon_{p_x p_y},
\end{split}
\end{equation}
where we have retained the $\mu = x$ or $y$ labels on the orbital states.

\subsection{Second-order Schrieffer-Wolff transform} \label{app:schrieffer_wolff}

Finally, we describe the second-order Schrieffer-Wolff (SW) transform used to compute the dynamics of the 6-level low-energy subspace.
Given all the single-particle integrals (see Sec.~\ref{app:random_field_generation}), we can compute matrix elements between any pair of basis states.
From these matrix elements, we construct the following block Hamiltonian:
\begin{equation}
    H_\mathrm{tot} 
    = \begin{pmatrix}
        H_\mathrm{g} & H_\mathrm{ge} \\ H_\mathrm{ge}^\dagger & H_\mathrm{e}
    \end{pmatrix}
\end{equation}
where $H_\mathrm{g}$ is the $ 6 \times 6$ low-energy Hamiltonian [Eq.~(\ref{eq:ham_first_order}) in the main text], $H_\mathrm{ge}$ includes all couplings between the low-energy subspace and the 32 excited orbital states described above, and
\begin{equation}
    H_\mathrm{e} = H_B + E_{\mathrm{orb},2e} + U_\mathrm{qw}
\end{equation}
describes the excited orbital states, including contributions from Zeeman energies ($H_B$), the orbital energy $E_{\mathrm{orb},2e}$, and the disorder-induced couplings described above.
Again, we ignore spin-orbit coupling in $H_\mathrm{e}$, which would only impact the low-energy dynamics at higher order.

To apply the SW transform, we then diagonalize $H_\mathrm{tot}$ using the block-diagonal rotation
\begin{equation}
    U_\mathrm{tot} = \begin{pmatrix}
        U_\mathrm{g} & 0 \\ 0 & U_\mathrm{e}
    \end{pmatrix}
\end{equation}
where $U_\mathrm{g} = U_v^{2e}$ in Eq.~(\ref{eq:urot_valley}) in the main text, and $U_\mathrm{e}$ is determined by diagonalizing $H_\mathrm{e}$.
The matrix elements and eigen-energies of $\tilde H_\mathrm{tot} = U_\mathrm{tot} H_\mathrm{tot} U_\mathrm{tot}^{-1}$ are used to compute the second order SW corrections given in Eq.~(\ref{eq:ham_second_order}).

We can see the results of our second-order Hamiltonian in Fig.~\ref{fig:energies}(b), where we plot the instantaneous eigen-energies of our low-lying orbital subspace across a region of heterostructure.
In this plot, the fluctuations of the valley landscape are obvious, shown in the $\ket{T_-^\mathrm{val}}$ and $\ket{T_+^\mathrm{val}}$ energies.
We have subtracted off a uniform $2 \epsilon_s$ from each energy, and the remaining fluctuations of the $\ket{S_0^\mathrm{val}}$ states reveal second-order shifts to the energy.

\section{Generating random valley coupling fields} \label{app:random_field_generation}

In this section, we describe how we generate random valley coupling fields for use in shuttling simulations.
These random fields are listed in Eqs.~(\ref{eq:valley_preserving_ints_supp}) and (\ref{eq:valley_flipping_ints_supp}).
All of these terms depend on the form of the quantum well potential $U_\text{qw}$, which induces these valley-flipping and valley-preserving random fields.
We can model the quantum well potential as a sum of a deterministic (or ``virtual crystal'') component and a random component:
\begin{equation}
U_\text{qw} = U_\text{qw}^\text{vc} + \delta U_\text{qw}
\end{equation}
where $U_\text{qw}^\text{vc}$ is given in Eq.~(\ref{eq:U_qw_vc}), and 
\begin{equation}
    \delta U_\text{qw} = \Delta E_c \frac{\delta X_{jkl}}{1 - X_s}
\end{equation}
where $\delta X_{jkl} = X_{jkl} - \bar X_{jkl}$, $X_{jkl}$ is the Si concentration in cell at position $(x_j, y_k, z_l)$, and $\bar X_{jkl}$ is the expected Si concentration in this cell, as determined by the heterostructure profile.
Alloy disorder contributes to the random term, $\delta U_\text{qw}$, whereas step disorder in the quantum well interface modifies $U_\text{qw}^\text{vc}$.
As we justify in Appendix~\ref{app:steps}, we ignore interface steps in our simulations. 
In this regime, it suffices to replace $U_\text{qw}$ with $\delta U_\text{qw}$ in the field definitions, allowing us to derive simple statistical relationships between these fields.
Using these statistical relationships, we can randomly generate all relevant energies in our simulations.
An example of the six randomly-generated low-energy states relevant to two-electron shuttling is shown in Fig.~\ref{fig:energies}(b).

\begin{figure*}[t] 
	\includegraphics[width=13cm]{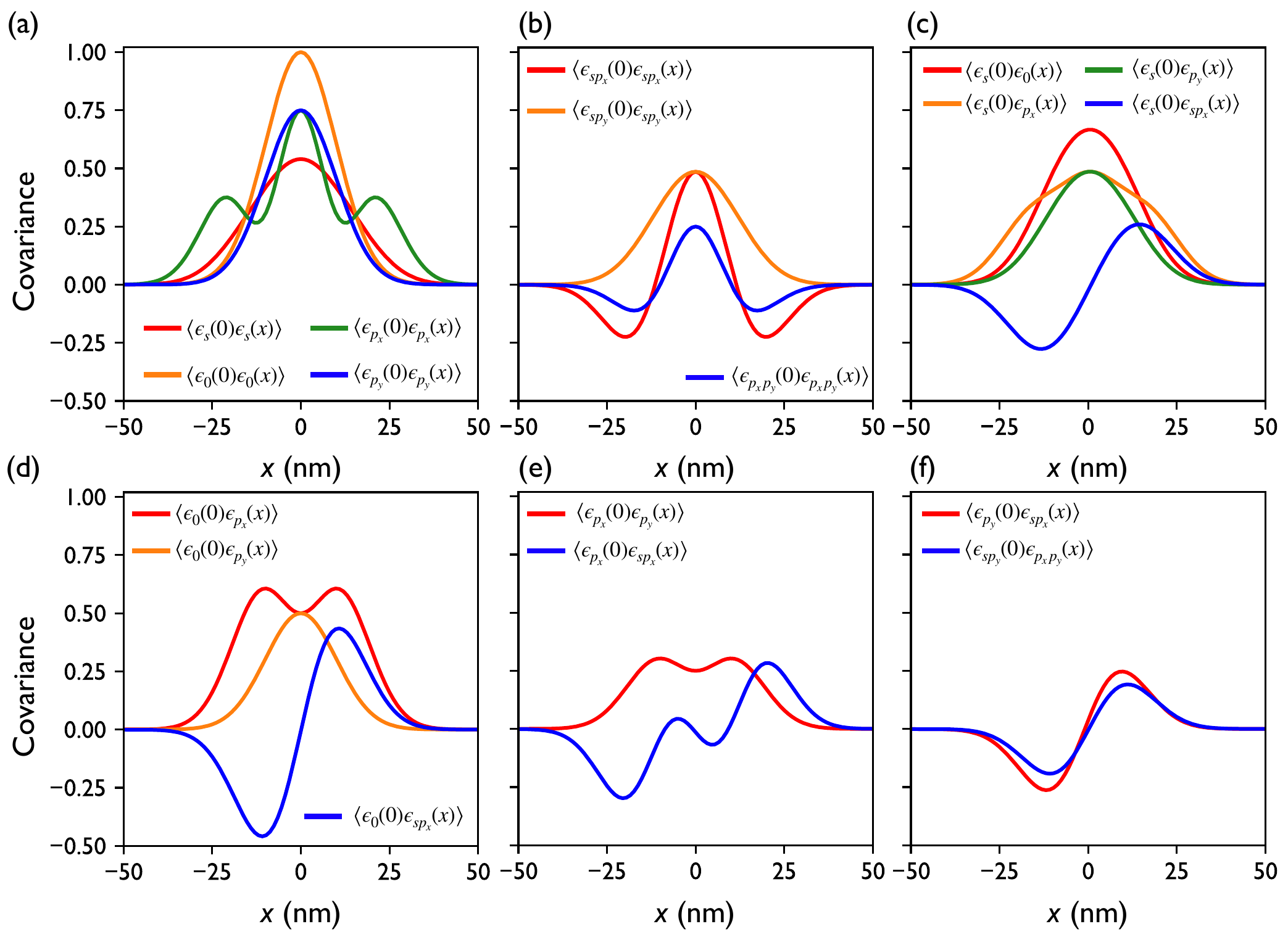}
	\centering
	\caption{Spatial covariance functions between the real fields used in shuttling simulations, in units for which $\sigma_{\Delta_0}^2 = 1$. (a-b) The autocorrelation functions for each of the seven real fields. (c-f) The nonzero cross-correlation functions, like those involving (c) $\epsilon_s$, (d) $\epsilon_0$, (e) $\epsilon_{p_x}$, and (f) $\epsilon_{p_y}$ and $\epsilon_{p_x p_y}$. Covariance functions are computed for $E_{\text{orb},1e} = 4$~\SI{}{\milli\electronvolt}.}
	\label{fig:cov_funcs}
\end{figure*}

For our two-electron shuttling simulations, we need to generate several valley coupling fields in a self-consistent manner.
We have both real and complex fields, which we assume are uncorrelated with each other.
We also assume the real and imaginary components of each complex field are uncorrelated.
Following Ref.~\cite{Woods:2024arXiv}, we discretize our shuttling segment into pieces of size \SI{1}{\nano\meter}, and we compute the covariance between each field at $x$, with each other field at $x'$.
For the real fields, this produces the covariance matrix
\begin{equation}
    \Sigma = 
    \begin{pmatrix}
        \Sigma_{\epsilon_s \epsilon_s} & 
        \Sigma_{\epsilon_s \epsilon_0} & 
        \Sigma_{\epsilon_s \epsilon_{p_x}} & 
        \Sigma_{\epsilon_s \epsilon_{p_y}} &  
        \Sigma_{\epsilon_s \epsilon_{sp_x}} & 
        \Sigma_{\epsilon_s \epsilon_{sp_y}} & 
        \Sigma_{\epsilon_s \epsilon_{p_x p_y}}
        \\
        \Sigma_{\epsilon_s \epsilon_0}^T & 
        \Sigma_{\epsilon_0 \epsilon_0} & 
        \Sigma_{\epsilon_0 \epsilon_{p_x}} & 
        \Sigma_{\epsilon_0 \epsilon_{p_y}} & 
        \Sigma_{\epsilon_0 \epsilon_{sp_x}} & 
        \Sigma_{\epsilon_0 \epsilon_{sp_y}} & 
        \Sigma_{\epsilon_0 \epsilon_{p_x p_y}} 
        \\
        \Sigma_{\epsilon_s \epsilon_{p_x}}^T & 
        \Sigma_{\epsilon_0 \epsilon_{p_x}}^T & 
        \Sigma_{\epsilon_{p_x} \epsilon_{p_x}} & 
        \Sigma_{\epsilon_{p_x} \epsilon_{p_y}} & 
        \Sigma_{\epsilon_{p_x} \epsilon_{sp_x}} & 
        \Sigma_{\epsilon_{p_x} \epsilon_{sp_y}} & 
        \Sigma_{\epsilon_{p_x} \epsilon_{p_x p_y}}
        \\
        \Sigma_{\epsilon_s \epsilon_{p_y}}^T & 
        \Sigma_{\epsilon_0 \epsilon_{p_y}}^T & 
        \Sigma_{\epsilon_{p_x} \epsilon_{p_y}}^T &
        \Sigma_{\epsilon_{p_y} \epsilon_{p_y}} & 
        \Sigma_{\epsilon_{p_y} \epsilon_{sp_x}} & 
        \Sigma_{\epsilon_{p_y} \epsilon_{sp_y}} & 
        \Sigma_{\epsilon_y \epsilon_{p_x p_y}}
        \\
        \Sigma_{\epsilon_s \epsilon_{sp_x}}^T & 
        \Sigma_{\epsilon_0 \epsilon_{sp_x}}^T & 
        \Sigma_{\epsilon_{p_x} \epsilon_{sp_x}}^T & 
        \Sigma_{\epsilon_{p_y} \epsilon_{sp_x}}^T & 
        \Sigma_{\epsilon_{sp_x} \epsilon_{sp_x}} & 
        \Sigma_{\epsilon_{sp_x} \epsilon_{sp_y}} & 
        \Sigma_{\epsilon_{sp_x} \epsilon_{p_x p_y}}
        \\
        \Sigma_{\epsilon_s \epsilon_{sp_y}}^T & 
        \Sigma_{\epsilon_0 \epsilon_{sp_y}}^T & 
        \Sigma_{\epsilon_{p_x} \epsilon_{sp_y}}^T & 
        \Sigma_{\epsilon_{p_y} \epsilon_{sp_y}}^T & 
        \Sigma_{\epsilon_{sp_x} \epsilon_{sp_y}}^T & 
        \Sigma_{\epsilon_{sp_y} \epsilon_{sp_y}} & 
        \Sigma_{\epsilon_{sp_y} \epsilon_{p_x p_y}}
        \\
        \Sigma_{\epsilon_{s} \epsilon_{p_x p_y}}^T & 
        \Sigma_{\epsilon_{0} \epsilon_{p_x p_y}}^T & 
        \Sigma_{\epsilon_{p_x} \epsilon_{p_x p_y}}^T & 
        \Sigma_{\epsilon_{p_y} \epsilon_{p_x p_y}}^T & 
        \Sigma_{\epsilon_{sp_x} \epsilon_{p_x p_y}}^T & 
        \Sigma_{\epsilon_{sp_y} \epsilon_{p_x p_y}}^T & 
        \Sigma_{\epsilon_{p_x p_y} \epsilon_{p_x p_y}}
    \end{pmatrix} .
\end{equation}

Each of the $\Sigma_{\alpha \beta}$, for fields $\alpha$ and $\beta$, is an $N \times N$ covariance matrix, where $N$ is the number of discretized points in the shuttling segment (at intervals of \SI{1}{\nano\meter}).
These matrices are constructed such that $\Sigma_{\alpha \beta}^{ij} = \langle \alpha(x_i) \beta(x_j)\rangle$, where $\langle \cdot \rangle$ is the expectation value, and $x_i$ is a position along the shuttling segment.
The covariance matrices used to generate the real and imaginary components of the complex-valued fields $\Delta_\alpha$ are $\Sigma_{\mathrm{Re}\Delta_\alpha \mathrm{Re} \Delta_\beta} = \Sigma_{\mathrm{Im}\Delta_\alpha \mathrm{Im} \Delta_\beta} = \frac{1}{2} \Sigma_{\epsilon_\alpha \epsilon_\beta}$.
The factor of $\frac{1}{2}$ comes from the division of the variance into the real and imaginary components, $\langle \alpha^* \beta \rangle = \langle \Re \alpha \Re \beta \rangle + \langle \Im \alpha \Im \beta \rangle$, for independent, zero-mean real and imaginary components.

To define these covariance matrices, we need covariance functions between the different fields, $\langle \epsilon_\alpha(x) \epsilon_\beta(x')\rangle = \langle \epsilon_\alpha(0) \epsilon_\beta(x' - x) \rangle$.
Some of these can be computed theoretically.
Direct calculations yield
\begin{equation} \label{eq:analytic_cov_funcs}
\begin{split}
    \langle \epsilon_0(x) \epsilon_0 (x')\rangle &= \sigma_{\Delta_0}^2 \exp \left( - \frac{(x'-x)^2}{2 a_\text{dot}^2}\right) 
    \\
    \langle \epsilon_{p_x}(x) \epsilon_{p_x} (x')\rangle &= \frac{\sigma_{\Delta_0}^2}{4} \left[3 - \frac{2(x'-x)^2}{a_\text{dot}^2} + \frac{(x'-x)^4}{a_\text{dot}^4} \right] \exp \left( - \frac{(x'-x)^2}{2 a_\text{dot}^2}\right)
    \\
    \langle \epsilon_{p_y}(x) \epsilon_{p_y} (x')\rangle &= \frac{3 \sigma_{\Delta_0}^2}{4} \exp \left( - \frac{(x'-x)^2}{2 a_\text{dot}^2}\right)
    \\
    \langle \epsilon_0(x) \epsilon_{p_x} (x')\rangle &= \frac{\sigma_{\Delta_0}^2}{2} \left[ 1 + \frac{(x'-x)^2}{a_\text{dot}^2} \right]  \exp \left( - \frac{(x'-x)^2}{2 a_\text{dot}^2}\right)
    \\
    \langle \epsilon_0(x) \epsilon_{p_y} (x')\rangle &= \frac{\sigma_{\Delta_0}^2}{2} \exp \left( - \frac{(x'-x)^2}{2 a_\text{dot}^2}\right)
    \\
    \langle \epsilon_{p_x}(x) \epsilon_{p_y} (x')\rangle &= \frac{\sigma_{\Delta_0}^2}{4} \left[ 1 + \frac{(x'-x)^2}{a_\text{dot}^2} \right]  \exp \left( - \frac{(x'-x)^2}{2 a_\text{dot}^2}\right)
    \\
    \langle \epsilon_{p_x p_y}(x) \epsilon_{p_x p_y}(x') \rangle & = \frac{\sigma_{\Delta_0}^2}{4} \left[1 - \frac{(x'-x)^2}{a_\mathrm{dot}^2} \right] \exp \left( -\frac{(x' - x)^2}{2 a_\mathrm{dot}^2}\right) .
\end{split}
\end{equation}

Note that, since we consider shuttling along $x$, the $y$-coordinate of all fields remains constant, simplifying some of the covariance relationships.
For completeness, we include details of the computation of $\langle \epsilon_0(x) \epsilon_0(x')\rangle$; the other covariance functions in Eq.~(\ref{eq:analytic_cov_funcs}) are determined in a similar fashion.
We can express the spatial covariance of $\epsilon_0$ as
\begin{equation}
    \langle \epsilon_0(x) \epsilon_0(x')\rangle 
    = \langle \int d\mathbf{r}'' \int d\mathbf{r}''' \;  \psi_0^2 (x'' - x,y'') \psi_0^2(x''' - x', y''') \psi_z^2(z'') \psi_z^2(z''') \delta U_\text{qw}(\mathbf{r}'') \delta U_\text{qw} (\mathbf{r}''') \rangle
\end{equation}
We approximate each integral as a discrete sum, using the transformation
\begin{equation} \label{eq:int_to_sum}
    \int d \mathbf{r} \rightarrow \Delta x \Delta y \Delta z \sum_{x, y, z},
\end{equation}
where the sum is over discrete, rectangular cells, as described in Appendix~\ref{app:fci}.
Using Eq.~(\ref{eq:int_to_sum}), we obtain
\begin{equation} \label{eq:cov_calc_partial}
    \langle \epsilon_0(x) \epsilon_0(x')\rangle = \left( \Delta x \Delta y \Delta z\right)^2 \sum_{x'',y'',z''} \sum_{x''', y''', z'''} \psi_0^2 (x''-x,y'') \psi_0^2(x''' - x', y''') \psi_z^2(z'') \psi_z^2(z''')  \langle \delta U_\text{qw}(\mathbf{r}'') \delta U_\text{qw} (\mathbf{r}''') \rangle
\end{equation}
Since $\delta U_\text{qw}$ is zero-mean and independent for different $\mathbf{r}$, $\langle \delta U_\text{qw}(\mathbf{r}'') \delta U_\text{qw} (\mathbf{r}''') \rangle = 0$ for $\mathbf{r}'' \neq \mathbf{r}'''$.
Furthermore, 
\begin{equation}
    \langle U_\text{qw}^2(\mathbf{r})\rangle = \frac{1}{n_c} \left( \frac{\Delta E_c}{X_w - X_s} \right)^2 \bar X_{\mathbf{r}} (1 - \bar X_{\mathbf{r}}),
\end{equation}
where $\bar X_\mathbf{r}$ is the expected Si concentration at position $\mathbf{r}$, derived by taking the variance of the binomial distribution of Eq.~(\ref{eq:binom}).
So, we can simplify Eq.~(\ref{eq:cov_calc_partial}), resulting in 
\begin{multline} \label{eq:cov_calc_partial_2}
    \langle \epsilon_0(x) \epsilon_0(x')\rangle = \left( \Delta x \Delta y \Delta z\right)^2 \sum_{x'',y'',z''} \psi_0^2 (x''-x,y'') \psi_0^2(x''-x', y'') \psi_z^4(z'') \langle \delta U_\text{qw}^2(\mathbf{r}'') \rangle \\
    = \frac{\eta}{16} \frac{\Delta x \Delta y}{n_c} \left( \frac{\Delta E_c}{X_w - X_s} \right)^2 \int dx'' dy'' \; \psi_0^2(x''-x, y'') \psi_0^2(x''-x', y'') 
\end{multline}
where we have converted the sums over $x$ and $y$ back to integrals.
We have defined
\begin{equation} 
    \eta = a_0^2 \sum_z \psi_z^4(z) \bar X_z ( 1 - \bar X_z)
\end{equation}
and we have assumed $\Delta z = a_0 / 4$, which is the natural lattice spacing in the $z$ direction, and that the expected Si concentration $\bar X_z$ is only a function of $z$ (i.e. that there are no interface steps).
The ratio $\Delta x \Delta y / n_c = a_0^2 / 2$, since there are 2 atoms per unit cell per atomic layer in Si/SiGe.
Evaluating the integral in Eq.~(\ref{eq:cov_calc_partial_2}), we obtain the covariance relation reported in Eq.~(\ref{eq:analytic_cov_funcs}), where we note that \cite{Losert:2023p125405}
\begin{equation} \label{eq:sigma_delta_0}
    \sigma_{\Delta_0}^2 = \frac{\eta}{\pi} \left[ \frac{a_0 \Delta E_c}{8 a_\text{dot} (X_w - X_s)} \right]^2.
\end{equation}

Other covariance functions can be set to zero on symmetry grounds:
\begin{equation}
    \begin{split}
        \langle \epsilon_s(x) \epsilon_{sp_y}(x')\rangle &= 0 
        \\
        \langle \epsilon_0(x) \epsilon_{sp_y}(x')\rangle &= 0 
        \\
        \langle \epsilon_{p_x}(x) \epsilon_{sp_y}(x')\rangle &= 0 
        \\
        \langle \epsilon_{sp_x}(x) \epsilon_{sp_y}(x')\rangle &= 0 
        \\
        \langle \epsilon_{p_y}(x) \epsilon_{sp_y}(x')\rangle &= 0 
        \\
        \langle \epsilon_{s p_x}(x) \epsilon_{sp_y}(x')\rangle &= 0 
        \\
        \langle \epsilon_s(x) \epsilon_{p_x p_y}(x')\rangle & = 0
        \\
        \langle \epsilon_0(x) \epsilon_{p_x p_y}(x')\rangle & = 0
        \\
        \langle \epsilon_{p_x}(x) \epsilon_{p_x p_y}(x')\rangle & = 0
        \\
        \langle \epsilon_{p_y}(x) \epsilon_{p_x p_y}(x')\rangle & = 0
        \\
        \langle \epsilon_{s p_x}(x) \epsilon_{p_x p_y}(x')\rangle & = 0.
    \end{split}
\end{equation}
The remaining covariance relationships involving $\epsilon_s$ must be computed numerically, which can be done using the envelope functions $\psi_s$, $\psi_0$, and $\psi_{p_x (p_y)}$.
The envelope function $\psi_s$ is obtained through EM FCI simulations, detailed above. 
(Since we compute these integrals to first order in perturbation theory, we can safely ignore disorder and valley coupling when we determine envelope functions.)
The $x$ and $y$ components of $\psi_0$ and $\psi_{p_x (p_y)}$ are given in Eq.~(\ref{eq:SHO}); we assume these wavefunctions share the same $z$ component as $\psi_s$.
For example, to compute the covariance between $\epsilon_s$ and $\epsilon_0$, we perform the following calculation:
\begin{multline} \label{eq:example_cov_calc}
    \langle \epsilon_s(x) \epsilon_0(x') \rangle = \int d\mathbf{r}'' \; \psi_s^2(x'' - x, y'') \psi_0^2(x'' - x', y'') \langle \delta U_\text{qw}^2 \rangle = \xi \int dx'' dy'' \psi_s^2(x'' - x, y'') \psi_0^2(x'' - x', y'') \\
    = \sigma_{\Delta_{s0}}^2  \frac{\int dx'' dy'' \; \psi_s^2(x'' - x,y'') \psi_0^2(x'' - x',y'')}{\int dx \; dy \; \psi_s^2(x,y) \psi_0^2(x,y)} ,
\end{multline}
where $\xi = \int dz \; \psi_z^4(z) \langle \delta U_\text{qw}^2 (z) \rangle$.
The integrals over $z$ factor out, since we do not include interface steps.
As a result, $\psi$ are separable into lateral and vertical components, and $\langle \delta U_\text{qw}^2 \rangle$ is a function only of $z$.
Thus, if we can provide $\sigma_{\Delta_{s0}^2} = \langle \epsilon_{s0}^2 \rangle = \langle |\Delta_{s0}|^2 \rangle$, we need only to compute the wavefunction overlaps on the right-hand side of Eq.~(\ref{eq:example_cov_calc}), which form the correlation function.
We can estimate $\sigma_{\Delta_{s0}}^2$ by relating it to $\sigma_{\Delta_0}^2$, which we take as a free parameter in this work:
\begin{equation}
    \sigma_{\Delta_{s0}}^2 = \sigma_{\Delta_0}^2 \frac{\int d\mathbf{r} \; \psi_s^2(\mathbf{r}) \psi_0^2(\mathbf{r})}{\int d\mathbf{r} \; \psi_0^4(\mathbf{r})} .
\end{equation}
Thus, through these overlap integrals, we can compute all covariance relations numerically.
These covariance relationships are illustrated in Fig.~\ref{fig:cov_funcs}.
All numerical covariance relationships are computed for $F_z = 1$ and \SI{10}{\milli\volt\per\nano\meter}, which are averaged to produce the covariances shown in Fig.~\ref{fig:cov_funcs}.

With these covariance matrices, we can generate random instantiations of the fields in question, following the methods of Woods et al.~\cite{Woods:2024arXiv}.
First, we diagonalize $\Sigma$ through the rotation $\Sigma' = U \Sigma U^\dag$, where $\Sigma'$ is a diagonal matrix. 
Now, we generate $M$ random Gaussian variables, stored in vector $\mathbf{v}'$, using covariance $\Sigma'$, where $M = 7N$.
Transforming $\mathbf{v}'$ back to the original basis, $\mathbf{v} = U^\dag \mathbf{v}'$, we have the fields stored in $\mathbf{v}$ with the correct covariance structure.

To generate random fields over \SI{5}{\micro\meter} is computationally demanding.
To reduce the computational overhead, we instead generate segments of length $l = 1.2$~\SI{}{\micro\meter}.
Then, we interpolate between these segments at their ends, with an overlap distance $l_\mathrm{int} = 200$~\SI{}{\nano\meter}, which should be much larger than the dot correlation length $a_\text{dot}$ and much smaller than the overall segment length $l$.
We do not use standard linear interpolation between these segments, which does not preserve the variance of the combined field. 
Instead, in the overlap region, we use the following interpolation scheme for random field $\alpha$, where $\alpha_0$ and $\alpha_1$ are independently-generated segments of length $l$:
\begin{equation}
\alpha(x) = \sqrt{1 - \left( \frac{x - (l - l_\text{int})}{l_\text{int}}\right)^2} \alpha_0(x ) + \left(\frac{x-(l-l_\text{int})}{l_\text{int}}\right) \alpha_{1}(x-(l-l_\text{int})),
\end{equation}
where $x$ ranges from $l - l_\mathrm{int}$ to $l$.
While non-standard, the interpolation form used above preserves the total variance of $\alpha$, an important quantity in our simulations.
A linear interpolation, for example, would reduce the variance of the fields, resulting in smaller valley splittings and second-order couplings, inflating the shuttling fidelities.

\section{Comparing the valley splitting and the singlet-triplet splitting} 
\label{app:valley_st}

\begin{figure*}[t] 
	\includegraphics[width=16cm]{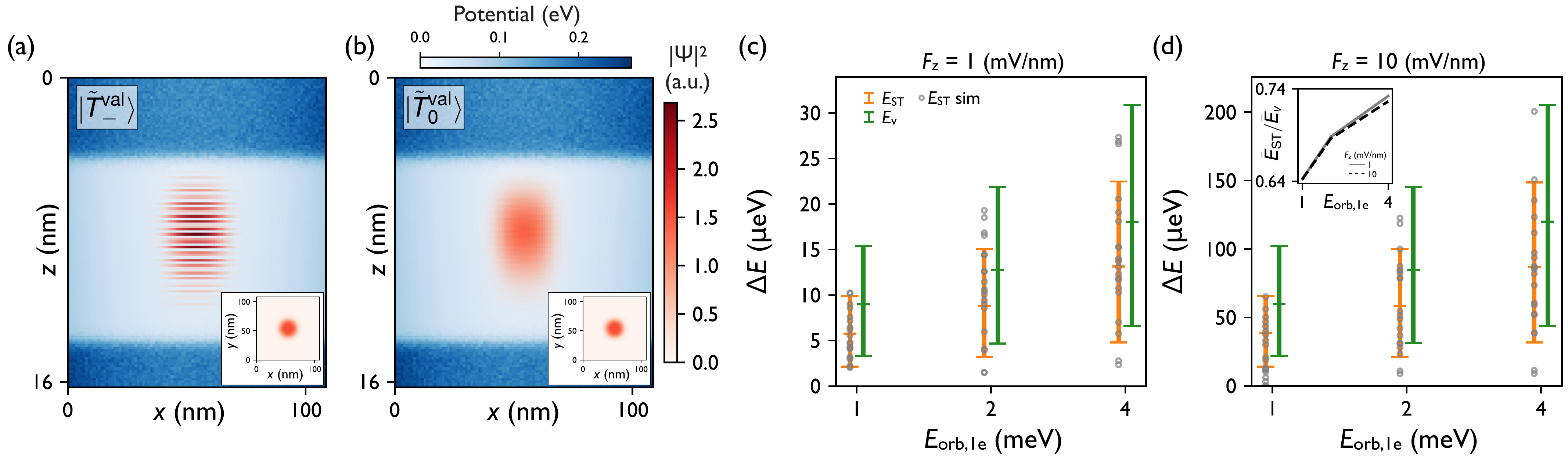}
	\centering
	\caption{The singlet-triplet splitting in isotropic two-electron quantum dots is well-explained by effective mass theory.
    (a) Ground state and (b) first excited state electron densities $|\Psi(x,z)|^2$ of a sample TB FCI simulation, where we have averaged over the $y$-dimension of the full 3D wavefunction. We use a lateral confinement $E_{\text{orb},1e} = 4$~\SI{}{\milli\electronvolt} and vertical field $F_z = 1$~\SI{}{\milli\volt\per\nano\meter}. 
    Insets show the same electron densities averaged over the $z$-dimension, $|\Psi(x,y)|^2$. 
    For illustrative purposes, we also show the potential used in these simulations (blue color bar), averaged along the $y$-dimension. 
    The fast oscillations of (a) are not visible in (b) because the former includes two valley ground states, while the latter includes a ground and an excited valley state. 
    (c) Mean and 10-90 percentile ranges for the single-electron valley splitting (green) and the singlet-triplet splitting (orange) for varying lateral confinement energies $E_{\text{orb},1e}$ and a vertical field $F_z = 1$~\SI{}{\milli\volt\per\nano\meter}. 
    Also included are 20 TB FCI simulations of $E_{ST}$ computed for the same heterostructure, where each point represents a different random alloy configuration (gray circles). 
    (d) The same quantities as (c), obtained for vertical fields $F_z = 10$~\SI{}{\milli\volt\per\nano\meter}. 
    Insets show ratio $E_{ST} / E_v$ for $E_{\mathrm{orb},1e}$ from 1 to 4 \SI{}{\milli\electronvolt}, for $F_z = 1$ and \SI{10}{\milli\volt\per\nano\meter}.
    }
	\label{fig:ev_est}
\end{figure*}

\begin{figure}[t]
\centering
\begin{minipage}{0.6\linewidth}
  \includegraphics[width=\linewidth]{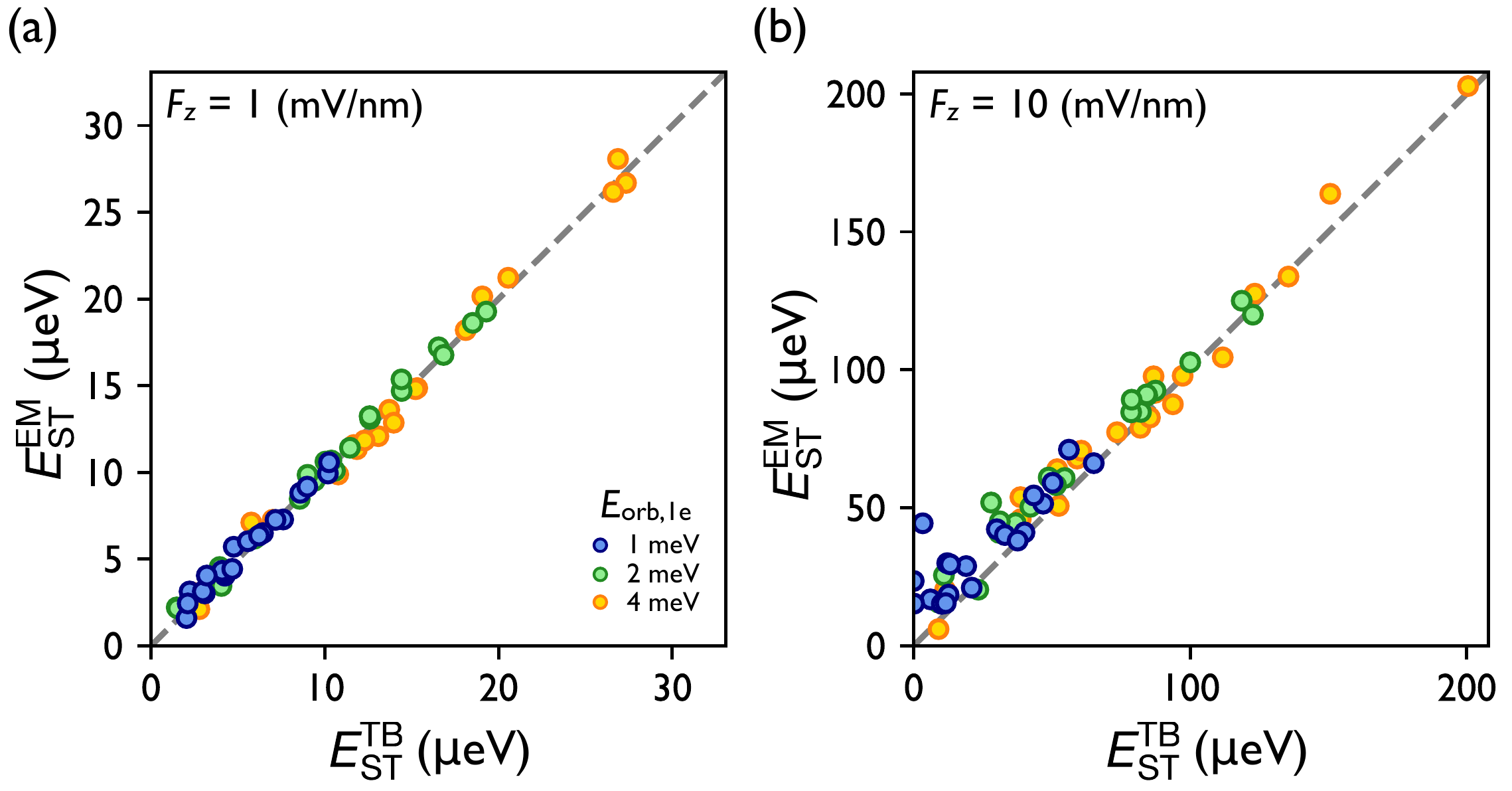}
\end{minipage}\hfill
\begin{minipage}{0.35\linewidth}
  \caption{\label{fig:em_tb_comparison}
    Comparing TB FCI simulations with effective mass theory. Here, we plot the singlet-triplet splitting extracted from TB FCI simulations of disordered quantum wells, $E_\text{ST}^\text{TB}$, with the singlet-triplet splittings computed numerically from effective mass theory, $E_\text{ST}^\text{EM}$, as described in Sec.~\ref{app:valley_st}. Each point represents a different instantiation of random alloy disorder. We include results for lateral confinement strengths $E_{\text{orb},1e} = $~\SI{1}{} (blue), \SI{2}{} (green), and \SI{4}{\milli\electronvolt} (orange), and for vertical fields (a) $F_z = 1$\SI{}{\milli\volt\per\nano\meter} and (b) \SI{10}{\milli\volt\per\nano\meter}.}
\end{minipage}
\end{figure}

In this section, we comment on the relationship between the valley splitting and the two-electron zero magnetic field singlet-triplet (ST) splitting.
The ST splitting is known to depend sensitively on the symmetry of the confinement potential \cite{Ercan:2021p235302, Ercan:2022p247701}.
For symmetric confinement potentials, it is bounded by the valley splitting.
On the other hand, for asymmetric potentials, electron-electron interactions cause the renormalization of the ST splitting, which is now of orbital character and often much smaller than the valley splitting.
Here, we consider only the symmetric case, so the ST splitting is well-described as a valley splitting.

To lowest order, the single-particle valley splitting is $E_v = 2|\Delta_0|$, where $\Delta_0$ is given in the main text Eq.~(\ref{eq:delta_main}).
Likewise, for two electrons, the (valley) ST splitting is given by $E_{ST} = 2|\Delta_s|$, where $\Delta_s$ is computed with Eq.~(\ref{eq:valley_flipping_ints_supp}).
At this order, the only difference between $E_v$ and $E_{ST}$ is the form of the ground state envelope function.
Thus, we can make a statistical comparison of these two quantities. 
From Ref.~\cite{Losert:2023p125405}, we know the average valley splitting $\bar E_v = \sqrt{\pi} \sigma_{\Delta_0}$ for the one-electron case, where $\sigma_{\Delta_0}^2 = \langle \Delta_0^2 \rangle$ is the variance of the single-electron $\Delta_0$. 
Likewise, in the two-electron case, using the same arguments, the average singlet-triplet splitting $\bar E_{ST} = \sqrt{\pi} \sigma_{\Delta_s}$, where $\sigma_{\Delta_s}^2 = \langle \Delta_s^2 \rangle$ is the variance of $\Delta_s$.

Using $\sigma_{\Delta_0}$ ($\sigma_{\Delta_s}$), we can characterize the distribution of resulting valley (singlet-triplet) splittings, which are given by the Rayleigh distribution \cite{Losert:2023p125405},
\begin{equation} \label{eq:rayleigh}
f_\text{Rayleigh}(z) = \frac{z^2}{\sigma^2} \exp \left[-\frac{z^2}{2 \sigma^2} \right] ,
\end{equation}
where the width parameter $\sigma = \sqrt{2} \sigma_{\Delta_{0(s)}}$.
We can also define the ratio of the mean ST splitting to the mean valley splitting for a given heterostructure,
\begin{equation} \label{eq:Ev_st_ratio}
	 \frac{\bar E_{ST}}{\bar E_v} = \frac{\sigma_{\Delta_s} }{ \sigma_{\Delta_0}} ,
\end{equation}
where $\bar E_v$ and $\bar E_{ST}$ are averaged over alloy disorder configurations.
Using the EM FCI simulations described in Section~\ref{app:fci}, we compute the single- and two-electron envelope functions for $E_{\text{orb},1e}$ from 1 to \SI{4}{\milli\electronvolt}, and for $F_z = 1$ and \SI{10}{\milli\volt\per\nano\meter}.
We then compute $\sigma_{\Delta_s}$ and $\sigma_{\Delta_0}$ from these envelope functions. 
The expected distributions of valley splitting (green bars) and ST splitting (orange bars) are shown as colored error bars in Fig.~\ref{fig:ev_est}(c) and (d).
In the inset to Fig.~\ref{fig:ev_est}(d), we plot the ratio $\bar E_{ST} / \bar E_v$, computed from Eq.~(\ref{eq:Ev_st_ratio}).
We notice that $E_{ST}$ is consistently smaller than $E_v$, where $\bar E_{ST} / \bar E_v $ is between 0.64 and 0.74, depending on the confinement strength.
This effect has a simple origin: since the two-electron envelope function is broadened by the Coulomb repulsion between electrons, the two-electron state is larger in area, averaging over more atomic disorder and therefore reducing $\sigma_{\Delta_s}$ compared to $\sigma_{\Delta_0}$.

We can also use the TB FCI simulations described in Section~\ref{app:fci} to confirm our statistical analysis of the ST splitting.
Using the same range of device parameters as above, we compute the ground state energy gap for 20 random instantiations of alloy disorder, plotted in Fig.~\ref{fig:ev_est}(c) and (d) [gray circles].
These data agree well with the statistical distributions of $E_{ST}$ computed above, validating our envelope function description of these systems.
For one of these simulations, we plot the ground state and first-excited state electron densities in Fig.~\ref{fig:ev_est}(a-b), where we have averaged the 3D wavefunctions over the $y$-dimension.
We see that the ground state has clear $|T_-^\text{val} \rangle$ character: since both electrons occupy the ground valley, the fast valley oscillations along $z$ are visible in the electron density. 
Since the first excited state contains both ground and excited valley components, which are 90$^\circ$ out of phase, the fast valley oscillations effectively average out.
Moreover, both ground and excited wavefunctions have clear $s$-like orbital character, as visible in the lateral electron densities shown in the insets to Figs.~\ref{fig:ev_est}(a-b), where we have instead averaged the 3D wavefunctions along the $z$-dimension.

Finally, we can directly compare our effective mass description of the ST splitting with tight-binding simulations, using the same 20 simulations in disordered quantum wells.
To do so, we compare the ST splitting computed from TB FCI, $E_\text{ST}^\text{TB}$, with the ST splitting computed from EM theory, $E_\text{ST}^\text{EM} = 2|\Delta_s|$.
We evaluate $\Delta_s$ numerically, using $\psi_s$ compute with EM FCI simulations, which completely ignore the valley degree of freedom.
In Fig.~\ref{fig:em_tb_comparison}, we plot the correlation between these quantities for electric fields (a) $F_z = 1$~\SI{}{\milli\volt\per\nano\meter} and (b) \SI{10}{\milli\volt\per\nano\meter}. 
Included in both plots are results for lateral confinement strengths $E_{\text{orb},1e} = $~\SI{1}{}, \SI{2}{}, and \SI{4}{\milli\electronvolt}.
In all cases the points lie near the line $y = x$ (gray dashed line), indicating that the first-order effective mass theory is capturing nearly all of the variation of $E_\text{ST}^\text{TB}$.
This agreement is particularly strong for $F_z = 1$~\SI{}{\milli\volt\per\nano\meter}, where vertical confinement is relatively weak, leading to small, disorder-dominated valley splittings -- the regime of interest for this work.

\section{Valley-orbit coupling due to interface steps} \label{app:steps}

\begin{figure}[t]
\centering
\begin{minipage}{0.6\linewidth}
  \includegraphics[width=\linewidth]{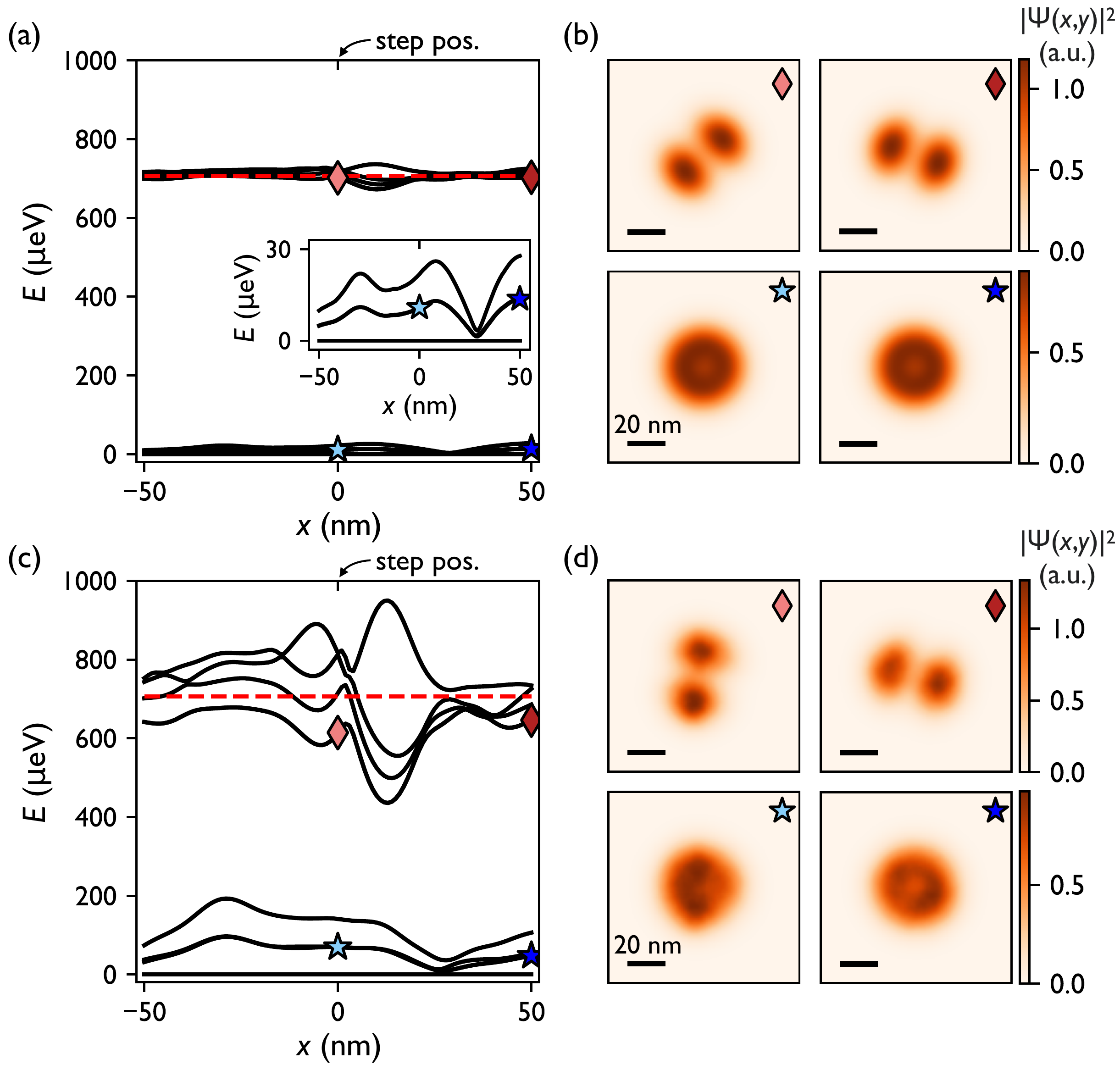}
\end{minipage}\hfill
\begin{minipage}{0.35\linewidth}
  \caption{\label{fig:step_energies}
    Low-energy spectrum as the dot moves across a single-monolayer interface step. (a) We plot several of the low-lying energies (in both the spin-singlet and spin-triplet subspaces) of a dot moving across an interface step, computed with TB FCI simulations. The orbital splitting, computed with EM FCI, is indicated as a red dashed line. Inset: the low-lying ground state orbital energies. We use a vertical field $F_z = 1$~\SI{}{\milli\volt\per\nano\meter}. (b) Sample electron densities for the states marked in (a) are plotted for dots positioned on the step and far from the step. (c) and (d) show the same data as (a) and (b), computed for $F_z = 10$~\SI{}{\milli\volt\per\nano\meter}.}
\end{minipage}
\end{figure}

In this section, we discuss the role of interface steps in the quantum well.
For single-electron quantum dots, single-monolayer steps in the quantum well interface have been explored as a source of valley splitting variability and valley-orbit coupling \cite{Kharche:2007p092109, Hosseinkhani:2020p043180, Ercan:2022p247701, Dodson:2022p146802, Tariq:2019p125309, Friesen:2006p202106, Goswami:2007p41, Culcer:2010p205315}.
More recently, it has been shown that steps are less important for the single-electron valley splitting in quantum wells without atomically sharp interfaces \cite{Losert:2023p125405}.
The same is true for inter-valley dipole moments \cite{Kanaar:2024arXiv}.
The picture is similar in the two-electron case.
We can treat the step as a perturbation:
\begin{equation} \label{eq:ham_step}
    U_\text{step} (x, z) = \frac{\Delta E_c}{1 - X_s} \delta X_l \Theta(x > x_\text{step}),
\end{equation}
where $\delta X_l = X_l - X_{l+1}$ and $\Theta$ is a Heaviside step function.
We can compute the inter-valley mixing induced by $U_\text{step}$ in a two-electron quantum dot.
For example, 
\begin{equation} \label{eq:intervalley_step}
    \langle T_-^\text{val} | U_\text{step} | T_0^\text{val} \rangle = \sqrt{2}\langle z_- | U_\text{step} | z_+ \rangle = \sqrt{2}\int d\mathbf{r} \; e^{-2 i k_0 z} \psi_s^2 U_\text{step} 
    = \frac{\sqrt{2}}{2} \frac{\Delta E_c a_0}{4 (1 - X_s)} \sum_l e^{-2 i k_0 z_l} \delta X_l \psi_s(z_l)^2
\end{equation}
where in the first step we have transformed the two-electron matrix element into single-particle integrals, and in the last step we have discretized the integral into a sum over cells using Eq.~(\ref{eq:int_to_sum}), and we have performed the $x$ and $y$ integrations. 
(We have assumed the step is at the center of the dot wavefunction, so the $x$-integration produces a factor of $\frac{1}{2}$.)
In Eq.~(\ref{eq:intervalley_step}), we note that the sum over layers $l$ includes a rapidly oscillating term, $e^{-2 i k_0 z_l}$. 
In the case of a perfectly sharp quantum well interface, $\delta X_l$ is nonzero only within one monolayer.
In this case, the rapidly oscillating term does not impact the sum, and the step strongly impacts the inter-valley coupling.
(This has been widely demonstrated in prior works, where interfaces have been assumed to be atomically sharp.)
However, in heterostructures with wider interfaces, $\delta X_l$ is nonzero across many monolayers, and the sum contains many nonzero terms.
Due to the rapidly oscillating summand, this sum will converge to zero.
Thus, for more realistic models, we do not expect steps to meaningfully contribute to inter-valley matrix elements.

While they do not meaningfully contribute to inter-valley matrix elements, steps do contribute to intra-valley terms.
For example, 
\begin{equation} 
    \langle T_-^\text{val} | U_\text{step} | T_-^\text{val} \rangle = 2 \langle z_- | U_\text{step} | z_- \rangle = 2 \int d\mathbf{r} \; \psi_s^2 H_\text{step} = \frac{\Delta E_c a_0}{4 (1 - X_s)} \sum_l \delta X_l \psi_s^2(z_l),
\end{equation}
In this case, we have no rapidly oscillating term.
So, as long as $\delta X_l$ and $\psi_s(z_l)$ have some overlap, the intra-valley contribution of a step is non-zero.
Moreover, if there is a large vertical field, $\psi_s$ is drawn closer to the top interface, increasing the product $\delta X_l \psi_s^2(z_l)$ and, therefore, increasing the intra-valley coupling.
While different intra-valley matrix elements will involve different envelope functions, this observation is generic. 

One way these intra-valley terms manifest is as diagonal corrections to the orbital energies.
In Fig.~\ref{fig:step_energies}, we plot the low-energy landscape for a two-electron quantum dot as it moves across a step positioned at $x = 0$, simulated with TB FCI methods, for a particular instantiation of random alloy disorder.
Results are shown for vertical fields (a-b) $F_z = 1$~\SI{}{\milli\volt\per\nano\meter} and (c-d) \SI{10}{\milli\volt\per\nano\meter}, using isotropic lateral confinement potentials of strength $E_{\mathrm{orb},1e} = 2.5$~\SI{}{\milli\electronvolt}.
We observe two separated manifolds, one corresponding to ground state orbital levels, and one to first-excited orbital states.
Electron densities for states in both manifolds, near the step and far away from it, are illustrated in Figs.~\ref{fig:step_energies}(b) and (d).
%We have plotted the two lowest states with spin configuration $|\downarrow \downarrow \rangle$ (corresponding to valley-orbit singlets), averaging the full 3D wavefunction over the $z$-dimension to obtain lateral densities $|\Psi(x,y)|^2$.
The orbital splitting for the (step-free) isotropic confinement potential, computed with EM FCI, is indicated as a red dashed line in (a) and (c).
In both cases, we observe fluctuations of the excited orbital energies near the position of the step (as well as fluctuations due to alloy disorder).
These fluctuations are weak for small vertical fields $F_z = 1$~\SI{}{\milli\volt\per\nano\meter}, but much larger for strong vertical fields $F_z = 10$~\SI{}{\milli\volt\per\nano\meter}.

Thus, in general, steps do contribute to intra-valley terms in the Hamiltonian, even for quantum well interfaces that are not atomically sharp.
However, in this work we are interested in the weak vertical field, small valley splitting limit, where these fluctuations are relatively small.
Moreover, realistic device modeling indicates the vertical field may be closer to $F_z = 1$~\SI{}{\milli\volt\per\nano\meter} in real devices \cite{Hollmann:2020p034068}.
Thus, for simplicity, we ignore interface steps in this work, focusing our simulations on landscapes determined exclusively by alloy disorder.

\section{The role of quantum dot anisotropy}

\begin{figure}[t]
\centering
\begin{minipage}{0.60\linewidth}
  \caption{\label{fig:anisotropic}
    Orbital energies of a two-electron quantum dot with $E_{\text{orb},1e}^y = 4$~\SI{}{\milli\electronvolt}, as we vary $E_{\text{orb},1e}^x$ from 3 to \SI{4}{\milli\electronvolt} (lower-left panel). We also show the corresponding ground state lateral electron densities $|\Psi_0(x,y)|^2$ obtained for $E_{\text{orb},1e}^x = 3$, 3.5, and \SI{4}{\milli\electronvolt} and $F_z = 1$~\SI{}{\milli\volt\per\nano\meter}, indicated with markers, as a demonstration of the Wigner physics.}
\end{minipage}\hfill
\begin{minipage}{0.35\linewidth}
  \includegraphics[width=\linewidth]{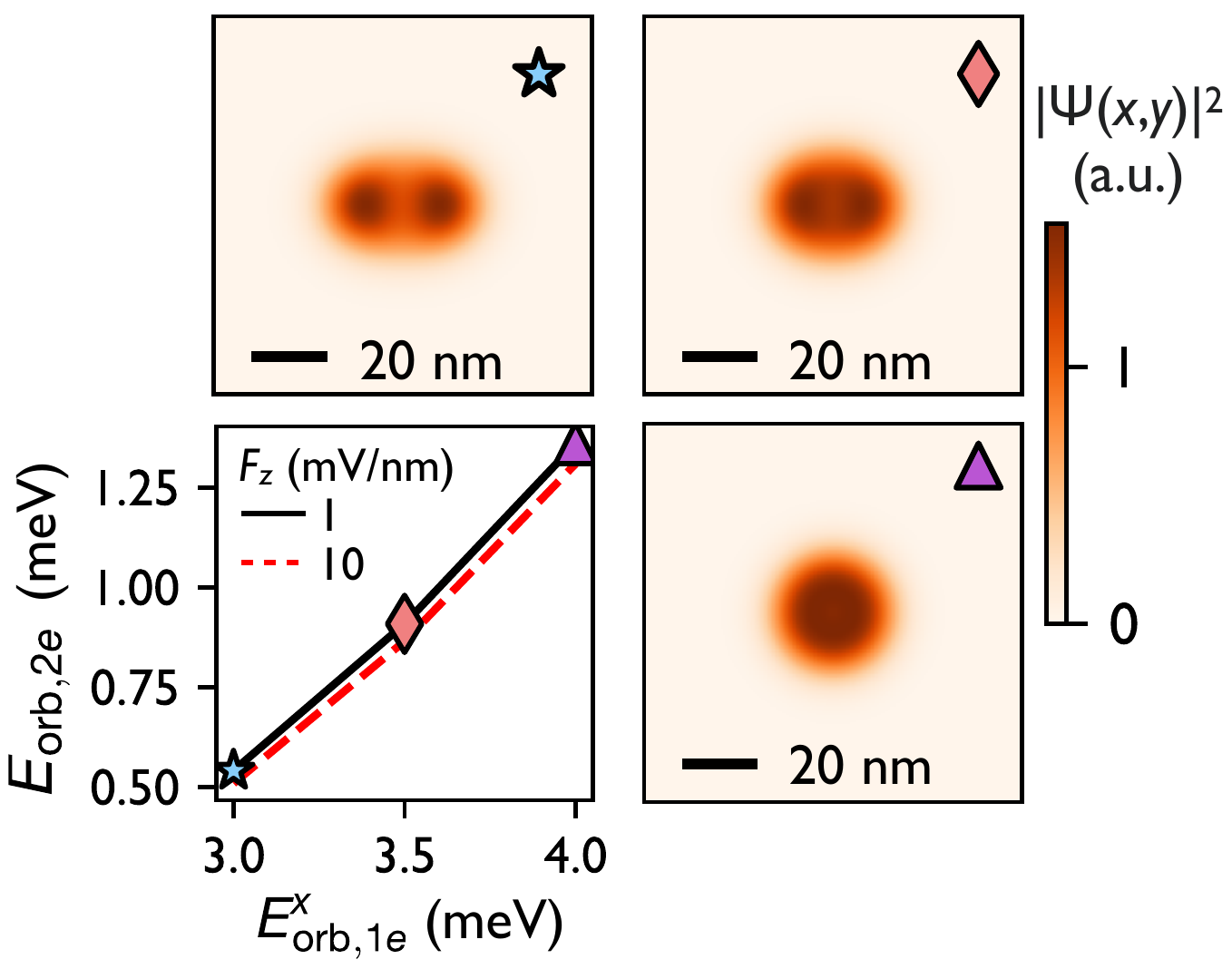}
\end{minipage}
\end{figure}

Here, we comment on the impact of quantum dot anisotropy on our shuttling scheme. 
It is well-known that in anisotropic two-electron quantum dots, the orbital splitting is significantly reduced due to the Coulomb interaction \cite{Ercan:2021p235302}.
In Fig.~\ref{fig:anisotropic}(a), we plot the lowest excited orbital energy for a quantum dot with $y$-confinement $E_{\text{orb},1e}^y = 4$~\SI{}{\milli\electronvolt} as we vary $E_{\text{orb},1e}^x$ from 3 to \SI{4}{\milli\electronvolt}.
We see that, for more anisotropic dots, the orbital spacing decreases rapidly -- a 25\% reduction in the confinement strength causes a $\sim 60$ \% reduction in $E_{\text{orb},2e}$.
Because second-order leakage terms scale like $E_{\text{orb},2e}^{-1}$, the resulting infidelity will scale like $E_{\text{orb},2e}^{-2}$, on average.
As discussed in the main text, this infidelity also depends on $k_{sp_\mu}$ matrix elements, which are likewise impacted by anisotropy.
By careful selection of the magnetic field orientation, it may be possible to mitigate these effects; a complete analysis of these dependencies is outside the scope of this work. 
Nonetheless, we expect best average performance for symmetrical dots.

\section{Dependence on spin-orbit coupling} \label{app:beta}

\begin{figure}[t]
\centering
\begin{minipage}{0.55\linewidth}
  \includegraphics[width=\linewidth]{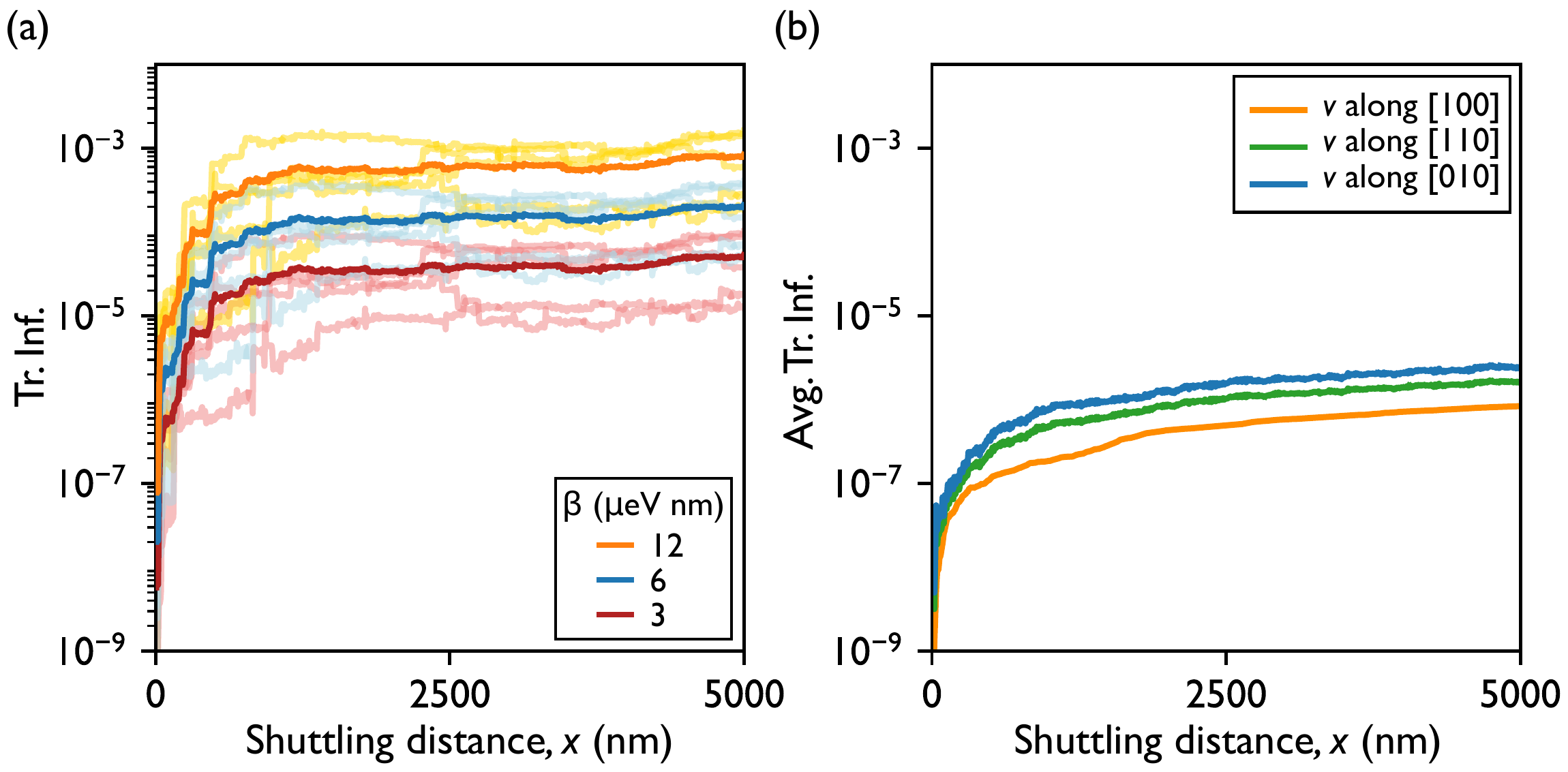}
\end{minipage}\hfill
\begin{minipage}{0.40\linewidth}
  \caption{\label{fig:so_strength}
    Exploring the role of SO coupling on shuttling fidelities.
    (a) Reducing the Dresselhaus spin-orbit coupling strength reduces the shuttling infidelity. Using the same five disorder landscapes, we compute the shuttling infidelity as described in the main text, using three values of the spin-orbit coupling parameter $\beta$. 
    (b) First-order spin-orbit coupling is not a dominant source of infidelity. Here, we show average infidelities across five random disorder landscapes, for shuttling along the [100] (orange), [110] (green), and [010] (blue) directions, where we have removed all second-order processes. We see that average infidelities are well below $10^{-4}$ in all cases.}
\end{minipage}
\end{figure}

In this section, we examine in more detail the role of SO coupling in our shuttling scheme.
First, we examine the impact of the SO coupling strength on second-order leakage.
For our shuttling scheme, leakage outside of the qubit subspace is proportional to the spin-orbit coupling strength, which is dominated by Dresselhaus-type interactions in Si.
In this work, we have used a Dresselhaus spin-orbit coupling strength $\beta = 12$~\SI{}{\micro\electronvolt\nano\meter}.
However, it is known that this parameter can be tuned by adjusting the vertical confinement strength of the dot \cite{Woods:2023p035418, Woods:2024arXiv, Nestoklon:2008p155328, Prada:2011p013009}.
In particular, $\beta$ is reduced in the low-$F_z$ limit.
In this section, we examine the performance of our shuttling scheme if we modulate $\beta$.
In Fig.~\ref{fig:so_strength}(a), we show shuttling infidelities as a function of distance for five instantiations of random alloy disorder, assuming dots with $E_{\text{orb},1e} = 4$~\SI{}{\milli\electronvolt}, $\sigma_{\Delta_0} = 20$~\SI{}{\micro\electronvolt}, and $v = 10$~\SI{}{\meter\per\second}.
The yellow data represent simulations with $\beta = 12$~\SI{}{\micro\electronvolt\nano\meter} [the same data is shown in Fig.~\ref{fig:fidelity}(g)].
Next, we reduce $\beta$ to \SI{6}{\micro\electronvolt\nano\meter} (blue) and \SI{3}{\micro\electronvolt\nano\meter} (red), simulating the system again with the same five random alloy landscapes.
Each time we reduce $\beta$ by a factor of 2, the total infidelity reduces by about a factor of 4.

Next, we evaluate the role of first-order spin-orbit coupling in our simulations.
In the main text, we claimed that this leakage is much weaker than second-order processes.
In Fig.~\ref{fig:so_strength}(b), we plot shuttling fidelities as a function of distance, where we have removed all second-order leakage terms from the simulations, isolating the leakage induced by first-order SOC.
We use $\beta = 12$~\SI{}{\micro\electronvolt\nano\meter} and the same valley and orbital configurations as in (a), setting $v = 10$~\SI{}{\meter\per\second} for shuttling in three directions: along [100] (orange), along [110] (green), and along [010] (blue). 
In all simulations, $B$ is oriented along [100].
Since $\psi_s$ is isotropic, we use the same 5 randomly generated valley coupling fields for each shuttling direction.
In all cases, we find infidelities much smaller than those reported in the main text.
Hence, regardless of shuttling direction, we expect second-order processes to dominate the shuttling infidelity.

\section{Preparing logical basis states}

\begin{figure*}[t] 
	\includegraphics[width=12cm]{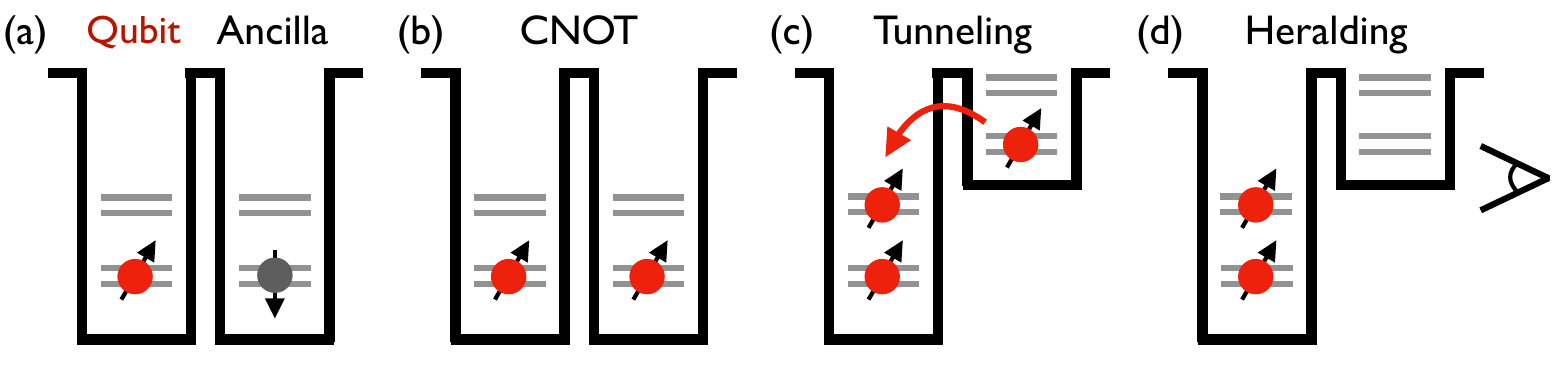}
	\centering
	\caption{Schematic illustration of a scheme to prepare the two-electron logical states from two Loss-DiVincenzo qubits. (a) The spin qubit in the left dot contains the quantum information we wish to shuttle, and the right dot contains an ancilla qubit initialized in a known state $|\downarrow \rangle$. (b) A CNOT operation, controlled on the left dot, entangles the two spins. Now, our qubit is a linear combination of $T_-^\text{spin}$ and $T_+^\text{spin}$. (c) The chemical potential in the right dot is lifted above the excited valley level of the left dot, allowing the right electron to tunnel into the left dot. (d) The tunneling process can be heralded by measuring the charge occupation of the right dot. }
	\label{fig:state_prep}
\end{figure*}

The shuttling scheme explored in this work relies on two-electron spin states of the form $\alpha \ket{\downarrow \downarrow} + \beta \ket{\uparrow \uparrow}$. 
This is an unusual basis for a spin qubit. 
Here, we briefly discuss one possible scheme to initialize these states from more typical single-spin Loss-Divincenzo qubits.
We start with a double quantum dot in the (1,1) charge configuration.
In the left dot, qubit $\ket{q_1} = \alpha \ket{\downarrow} + \beta \ket{\uparrow}$ contains the quantum information we wish to shuttle, and in the right dot, an ancilla qubit is initialized as $\ket{q_2} = \ket{\downarrow}$, as illustrated in Fig.~\ref{fig:state_prep}(a).
By performing a CNOT, controlled on $\ket{q_1}$, we prepare the state $\alpha \ket{\downarrow \downarrow } + \beta \ket{\uparrow \uparrow }$, illustrated in Fig.~\ref{fig:state_prep}(b).
This state has the correct spin configuration, but it is dispersed across a double dot.
By lifting the detuning in the right dot above the excited valley energy in the left dot, the electron in the right dot may tunnel into the left dot, as illustrated in Fig.~\ref{fig:state_prep}(c).
This tunneling only occurs if the valley phase in the left and right dots are not identical. 
However, in a valley landscape dominated by alloy disorder, this is almost surely the case.
Since the two electrons have identical spin, the only possible configuration for the two electrons in a single dot is a valley singlet.
The resulting state is $\alpha |S^\text{val} T_-^\text{spin}\rangle + \beta |S^\text{val} T_+^\text{spin} \rangle$.
Finally, the tunneling can be heralded by measuring the occupation in the right dot, as depicted in Fig.~\ref{fig:state_prep}(d).
The reverse process can be used to convert the encoded qubit back into a single-electron Loss-Divincenzo qubit, plus an ancilla.

\section{Achieving very low alloy disorder}

\begin{figure}[t]
\centering
\begin{minipage}{0.40\linewidth}
  \caption{\label{fig:low_disorder} 
    Very low average $\sigma_{\Delta_0}$ is achievable by tuning the vertical field.
    (a) Several vertical envelope functions $\psi_z$ are shown as we increase $F_z$ from 0 to \SI{10}{\milli\volt\per\nano\meter}. As the field increases, $\psi_z$ is pulled more strongly into the top interface.
    (b) Using $\psi_z$ from (a), we compute $\sigma_{\Delta_0}$ using Eq.~(\ref{eq:sigma_delta_0}) for $E_{\text{orb},1e} = 2$~\SI{}{\milli\electronvolt} (red), $E_{\text{orb},1e} = 4$~\SI{}{\milli\electronvolt} (orange), and $E_{\text{orb},1e} = 6$~\SI{}{\milli\electronvolt} (blue).}
\end{minipage}\hfill
\begin{minipage}{0.55\linewidth}
  \includegraphics[width=\linewidth]{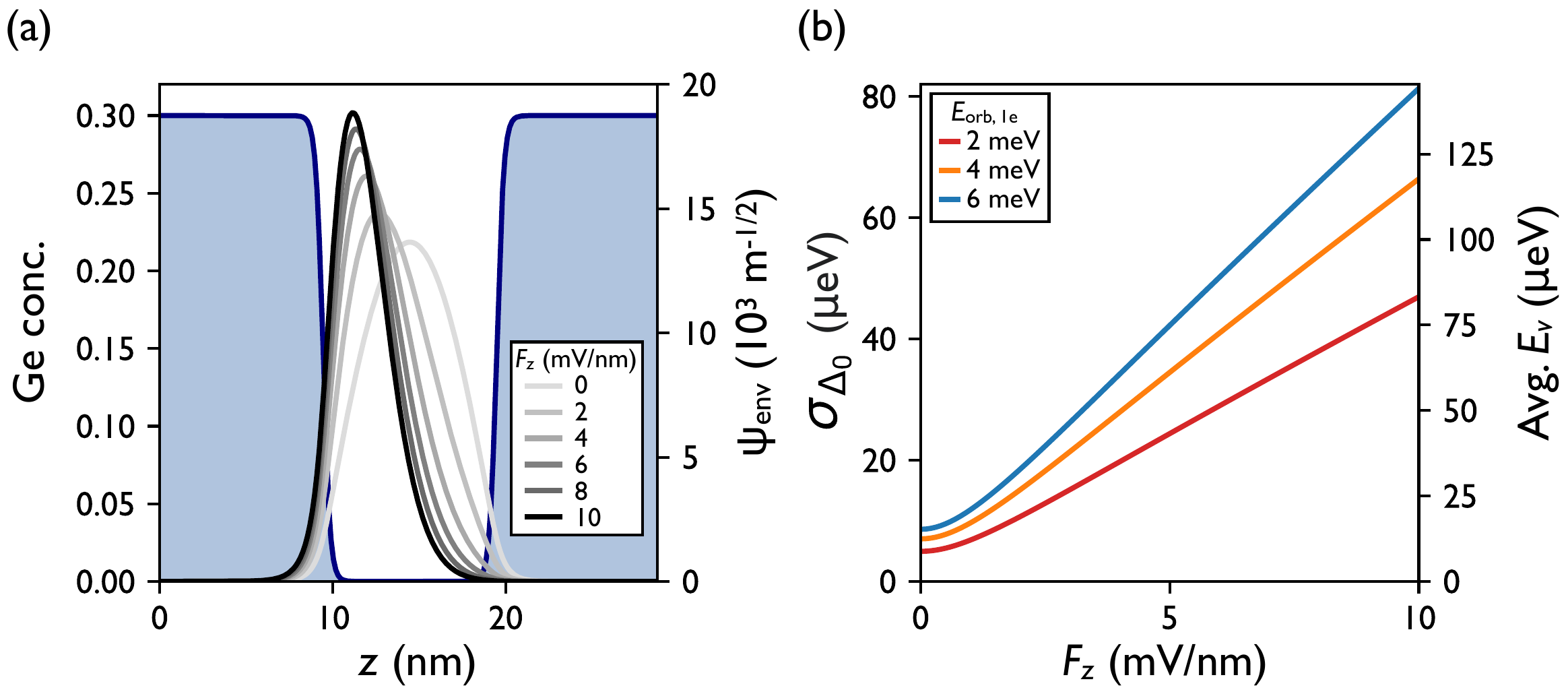}
\end{minipage}
\end{figure}

We have demonstrated in this work that spin shuttling fidelities are optimized when the alloy disorder, parameterized by $\sigma_{\Delta_0}$, is very low.
In this section, we comment on how to achieve such low-disorder devices.
For the device simulated in this work, we find that we can achieve $\sigma_{\Delta_0} \lesssim 10$~\SI{}{\micro\electronvolt} by adjusting the vertical field $F_z$, even for strongly confined quantum dots.
This puts our system within the low-disorder, large-confinement regime needed for high-fidelity shuttling.
In Fig.~\ref{fig:low_disorder}(a), we plot the quantum dot envelope function $\psi_z$ at different values of the vertical field $F_z$. 
At large vertical fields, the dot is strongly pushed into the top interface, causing the wavefunction to overlap with high-Ge layers and increasing $\sigma_{\Delta_0}$. 
In contrast, at low $F_z$, the wavefunction is not strongly confined to the top interface, and lower $\sigma_{\Delta_0}$ are possible.
In Fig.~\ref{fig:low_disorder}(b), we plot $\sigma_{\Delta_0}$ as a function of $F_z$ for three confinement strengths: $E_{\text{orb},1e} = 2$~\SI{}{\milli\electronvolt} (red), \SI{4}{\milli\electronvolt} (orange), and \SI{6}{\milli\electronvolt} (blue), where we compute $\sigma_{\Delta_0}$ using Eq.~(\ref{eq:sigma_delta_0}).
Thus, by modulating the vertical field, we can achieve the low values of $\sigma_{\Delta_0}$ required for high-fidelity shuttling. 
In addition, at these low field values, it is known that $\beta$ is significantly reduced, which further improves shuttling fidelity (see SM Section~\ref{app:beta}).
Hence, we view vertical field modulation as a promising strategy to achieve high-fidelity valley-free shuttling in Si.

\section{Estimating dephasing and relaxation rates}

In this section, we estimate decoherence rates for our logical spin qubits.
As well as $T_2$ dephasing rates, since our shuttling scheme utilizes excited spin-valley states, we estimate $T_1$ relaxation rates due to $1/f$ noise and electron-phonon coupling.
We demonstrate that leakage errors dominate both $T_1$ and $T_2$ processes in this system, and that the excited state lifetimes are much longer than the shuttling time and the dephasing time.

\subsection{Dephasing rates}
First, we consider dephasing of our spin qubit. 
We assume a qubit $T_2^* = 10$~\SI{}{\micro\second} and a noise correlation length $l_c = 100$~\SI{}{\nano\meter}.
From these, following Ref.~\cite{Losert:2024p040322}, we estimate the infidelity as
\begin{equation} \label{eq:dephasing_est}
    I_{\delta \phi} \approx \frac{l_c x}{(v T_2^*)^2}
\end{equation}
for shuttling distance $x$ and velocity $v$. 
The infidelity using Eq.~(\ref{eq:dephasing_est}) is plotted in Fig.~\ref{fig:fidelity}(g).

\subsection{Charge noise}
Next, we consider $1/f$ charge noise, ubiquitous in silicon quantum devices \cite{Wuetz:2023p1385, Connors:2022p940}.
For concreteness, we consider relaxation from $|0_L \rangle = |\tilde S^\text{val} T_-^\text{spin} \rangle$ to the ground state $|\tilde T_-^\text{val} S^\text{spin}\rangle$.
Relaxation rates from $|1_L\rangle$ will be nearly the same.
The charge-noise induced relaxation rate is given by \cite{Huang:2014p235315, Hosseinkhani:2021p085309}
\begin{equation}
    \Gamma_{1/f} = \frac{4 \pi e^2}{\hbar^2} S_E(\omega)|\mathbf{r}|^2
\end{equation}
where $S_E$ is the power spectral density of electric field fluctuations, $\omega$ energy difference between the ground and excited state, expressed as a frequency, $|\mathbf{r}| = |\langle \tilde T_-^\text{val} S^\text{spin}| (\hat x + \hat y) | 0_L \rangle|$ is the dipolar coupling between these states, and we have ignored $z$ due to the large vertical confinement energy in a quantum well.
Following Ref.~\cite{Hosseinkhani:2021p085309}, we relate electric field fluctuations to energy fluctuations via
\begin{equation}
    S_E(\omega) = \frac{S_\varepsilon(\omega)}{(e l_0)^2}
\end{equation}
where $l_0$ is the length scale between the dot and nearby charge fluctuators (we take $l_0 = 50$~\SI{}{\nano\meter}).
For $1/f$ noise, we use $S_\varepsilon(\omega) = S_0 / \omega$, where $S_0$ is the power spectral density at 1 Hz, which we set to \SI{1}{\micro\electronvolt^2\per\Hz} \cite{Hosseinkhani:2021p085309}.
Coupling of the form $\langle \tilde T_-^\text{val} S^\text{spin}| \mathbf{r} | 0_L \rangle$ is enabled by the spin-orbit interaction, which hybridizes $|0_L\rangle$ to states like $|T_0^\text{orb} \tilde T_-^\text{val} S^\text{spin} \rangle$ [see Eq.~(\ref{eq:excited_orbital_SO_couplings})].
This hybridization scales like $k_{sp} \beta / E_{\text{orb},2e}$, where $\beta$ represents either a Rashba or Dresselhaus SO coefficient.
The resulting orbital transition has dipolar matrix element $|\mathbf{r}_{01}| = |\langle T_-^\text{orb}| \mathbf{r} | T_0^\text{orb} \rangle |\sim |\langle \psi_{p_x} | \mathbf{r} | \psi_s \rangle|$, which we compute numerically for each $E_\text{orb, 1e}$ and is on the order of \SI{10}{\nano\meter}.
Thus, the dipolar operator $|\mathbf{r}| \sim \beta k_{sp_x} r_{01}/ E_{\text{orb},2e}$.
We can estimate the noise-induced relaxation rate as
\begin{equation}
    \Gamma_{1/f} \sim \frac{4 \pi}{\hbar l_0^2 E_{\text{orb},2e}^2} \big\vert r_{01} k_{sp_x} \beta \big\vert^2 \frac{1}{\Delta E_\text{ave}}
\end{equation}
where $\Delta E_\text{ave} \approx \bar E_v$ is the average splitting between the ground state and $|0_L\rangle$, approximately given by the valley splitting.
Setting $\bar E_v \approx \sqrt{\pi} \sigma_{\Delta_0}$ \cite{Losert:2023p125405} and using $k_{sp_x}$ and $E_{\text{orb},2e}$ computed for each $E_\text{orb,1e}$, we obtain $\Gamma_{1/f}$.
We estimate the resulting infidelity
\begin{equation}
    I_{1/f} \approx \Gamma_{1/f} t = \Gamma_{1/f} \frac{x}{v},
\end{equation}
plotted Fig.~\ref{fig:fidelity}.

\subsection{Phonon relaxation}
Lastly, we consider excited state relaxation due to phonon noise. 
In the electric dipole approximation, this relaxation rate is given by \cite{Yang:2013p3069, Tahan:2014p075302, Hosseinkhani:2021p085309}
\begin{equation} \label{eq:phonon_rate}
    \Gamma_\text{ph} = \frac{\Delta E^5 |\mathbf{r}|^2}{4 \pi \rho \hbar^6} \left( \frac{I_l}{v_l^7} + \frac{I_t}{v_t^7} \right)
\end{equation}
where $\rho = 2330$~\SI{}{\kilogram\per\meter^3} is the density of Si, $v_l = 9330$~\SI{}{\meter\per\second} and $v_t = 5420$~\SI{}{\meter\per\second} are the velocities of longitudinal and transverse modes, and the angular integrals are given by
\begin{equation}
\begin{split}
    I_l &= \frac{2}{35} \Xi_u^2 + \frac{4}{15}\Xi_u \Xi_d + \frac{2}{3}\Xi_d^2 \\
    I_t &= \frac{8}{105} \Xi_u^2
\end{split}
\end{equation}
for deformation potentials $\Xi_u = 8.77$~\SI{}{\electronvolt} and $\Xi_d = 5$~\SI{}{\electronvolt}.
Using the same estimate of $|\mathbf{r}|$ as above, and setting the energy gap $\Delta E \sim \bar E_v \sim \sqrt{\pi} \sigma_{\Delta_0}$, we estimate the phonon-induced decay rate.
The resulting infidelity is given by $I_{1/f} \approx \Gamma_{1/f} t$.

Of course, this rate depends sensitively on the average valley splitting, since there is a fifth-order dependence on $\Delta E$ in Eq.~(\ref{eq:phonon_rate}).
However, even using $\bar E_v = 100$~\SI{}{\micro\electronvolt} produces $\Gamma_\text{ph} \sim 3.5 \times 10^3$~\SI{}{\second^{-1}} for $E_\text{orb,1e} = 2$~\SI{}{\milli\electronvolt}, comparable with the leakage rate in our simulations and consistent with measurements of $T_1$ in a two-electron MOS device with large $E_v$ \cite{Yang:2013p3069}.
Since our shuttling scheme is most promising for $E_v < 50$~\SI{}{\micro\electronvolt}, we can safely ignore phonon relaxation.

\bibliography{bibliography}

\end{document}